\documentclass[twocolumn, twocolappendix]{aastex631}
\usepackage{enumitem}
\usepackage{xspace}
\usepackage{amsmath}
\usepackage{float}
\usepackage{booktabs}
\usepackage{natbib}

\shorttitle{Anthology of Fast Radio Burst Dispersion Measure Cross-Correlations}
\shortauthors{Sharma et al.}

\begin{document}


\title{Backlighting the Cosmic Web with Fast Radio Bursts: An Anthology of Dispersion Measure Cross-Correlations with Large-Scale Structure and Baryon Tracers}

\correspondingauthor{Kritti Sharma}
\email{kritti@caltech.edu}

\author[0000-0002-4477-3625]{Kritti Sharma}
\affiliation{Cahill Center for Astronomy and Astrophysics, MC 249-17 California Institute of Technology, Pasadena CA 91125, USA.}

\author[0000-0001-8356-2014]{Elisabeth Krause}
\affiliation{Department of Astronomy/Steward Observatory, University of Arizona, 933 North Cherry Avenue, Tucson, AZ 85721, USA}
\affiliation{Department of Physics, University of Arizona, 1118 E Fourth Street, Tucson, AZ 85721,
USA}

\author[0000-0002-7252-5485]{Vikram Ravi}
\affiliation{Cahill Center for Astronomy and Astrophysics, MC 249-17 California Institute of Technology, Pasadena CA 91125, USA.}
\affiliation{Owens Valley Radio Observatory, California Institute of Technology, Big Pine CA 93513, USA.}

\author[0000-0003-3312-909X]{Dhayaa Anbajagane}
\affiliation{Department of Astronomy and Astrophysics, University of Chicago, Chicago, IL 60637, USA}
\affiliation{Kavli Institute for Cosmological Physics, University of Chicago, Chicago, IL 60637, USA}

\author[0000-0002-7587-6352]{Liam Connor}
\affiliation{Center for Astrophysics, Harvard $\&$ Smithsonian, Cambridge, MA 02138-1516, USA.}

\author[0000-0001-5411-6920]{W.L. Kimmy Wu}
\affiliation{Cahill Center for Astronomy and Astrophysics, MC 249-17 California Institute of Technology, Pasadena CA 91125, USA.}

\author[0000-0003-4992-7854]{Simone Ferraro}
\affiliation{Lawrence Berkeley National Laboratory, 1 Cyclotron Road, Berkeley, CA 94720, USA}
\affiliation{Berkeley Center for Cosmological Physics, Department of Physics, University of California, Berkeley, CA 94720, USA}

\author[0000-0002-4577-8217]{Sebastian Grandis}
\affiliation{Universit\"at Innsbruck, Institut f\"ur Astro- und Teilchenphysik, Technikerstr. 25/8, 6020 Innsbruck, Austria}

\author[0000-0002-4598-9719]{David Alonso}
\affiliation{Department of Physics, University of Oxford, Denys Wilkinson Building, Keble Road, Oxford OX1 3RH, United Kingdom}

\author[0000-0001-6320-261X]{Yi-Kuan Chiang}
\affiliation{Academia Sinica Institute of Astronomy and Astrophysics (ASIAA), No. 1, Section 4, Roosevelt Road, Taipei 10617, Taiwan}

\author[0000-0002-4119-9963]{Casey J. Law}
\affiliation{Cahill Center for Astronomy and Astrophysics, MC 249-17 California Institute of Technology, Pasadena CA 91125, USA.}
\affiliation{Owens Valley Radio Observatory, California Institute of Technology, Big Pine CA 93513, USA.}

\author[0000-0003-3714-2574]{Pranjal R. S.}
\affiliation{Department of Astronomy/Steward Observatory, University of Arizona, 933 North Cherry Avenue, Tucson, AZ 85721, USA}

\author[0009-0008-5043-6220]{Samuel McCarty}
\affiliation{Center for Astrophysics, Harvard $\&$ Smithsonian, Cambridge, MA 02138-1516, USA.}

\author[0000-0001-5780-637X]{Shivam Pandey}
\affiliation{Department of Astronomy/Steward Observatory, University of Arizona, 933 North Cherry Avenue, Tucson, AZ 85721, USA}

\begin{abstract}

Fast Radio Bursts (FRBs) probe baryons permeating the cosmic web through their dispersion measures (DMs), which encode the integrated electron density along cosmological sightlines. Using 3,455 unique FRB sources from CHIME/FRB with $\sim 15$~arcmin localizations, we present an anthology of DM correlations with tracers of large-scale structure and baryonic matter at redshifts $z \lesssim 1.5$. We measure statistically significant correlations at $2.6-5\sigma$ with ten probes, including galaxies ($2.8\sigma$), weak gravitational lensing ($2.6\sigma$), cosmic infrared background ($4.0\sigma$), cosmic microwave background (CMB) lensing ($3.3\sigma$), thermal Sunyaev Zel'dovich (tSZ) effect ($3.8\sigma$), X-ray emission tracing galaxy clusters ($5.0\sigma$) and superclusters ($3.3\sigma$), soft X-ray background (SXRB, $4.1\sigma$), and radio continuum emission ($3.2\sigma$). These measurements reveal a consistent picture in which FRB sightlines intersecting overdense environments carry systematically larger DMs. Correlations with hot-gas tracers provide additional leverage on the strength of feedback, as they are strongly weighted towards the dense, bound gas. The measured amplitude of tSZ$\times$DM and SXRB$\times$DM correlations are consistent with theoretical predictions of baryon distribution from a DM-$z$ relation-inferred model with moderate feedback at $\sim 0.5\sigma$ level. Weaker feedback scenario is ruled out at $\sim 3.5\sigma$ by the SXRB$\times$DM correlation. Taken together, these measurements constitute a quantitative multi-tracer foundation for a new era in which FRBs from next generation facilities, such as BURSTT, CHORD, DSA, and SKA, in harmony with other probes, will map the baryon content of the full extent of the cosmic web.

\end{abstract}

\section{Introduction} \label{sec:introduction}

The distribution of baryons across cosmic scales encodes the cumulative imprint of galaxy formation processes, such as stellar winds~\citep{2018Galax...6..114Z}, supernovae~\citep{2009ApJ...695..292C}, active galactic nucleus~\citep[AGN;][]{2012ARA&A..50..455F}, and cosmic ray~\citep{2025OJAp....8E..66Q} feedback, that regulate how gas is retained, heated, and ejected from dark matter halos. Quantifying this distribution is not merely an astrophysical goal, it is a prerequisite for precision cosmology~\citep{2019OJAp....2E...4C}. The reach of next-generation weak gravitational lensing surveys~\citep{2011arXiv1110.3193L, 2019ApJ...873..111I} into the non-linear regime is fundamentally limited by uncertainties in baryonic feedback~\citep{2019OJAp....2E...4C}. Imposing conservative scale cuts~\citep{2025A&A...703A.158W, 2026arXiv260210065D} or marginalizing over flexible feedback models~\citep{2019JCAP...03..020S} inevitably sacrifices signal-to-noise, weakens the resulting cosmological constraints, and opens the way for parameter volume effects. Disentangling baryonic effects requires observational probes of the baryon distribution~\citep{2024MNRAS.534..655B, 2025arXiv250704476D, 2025PhRvD.112l3507H, 2025arXiv250607432P, 2025arXiv251202954S, 2026arXiv260417162S}.

The dispersion measure (DM) of Fast Radio Bursts~\citep[FRBs;][]{2022A&ARv..30....2P} encode the frequency-dependent dispersive delay proportional to the integrated free electron density along the line of sight as the pulse propagates though the intervening plasma. Formally, the observed DM includes contributions from the intergalactic medium (IGM) and the circumgalactic medium (CGM) of intervening halos (the cosmic component), along with the Galactic~\citep{2002astro.ph..7156C, 2003astro.ph..1598C}, and the FRB host galaxy contributions~\citep{2020ApJ...900..170Z, 2024ApJ...972L..26O, 2025A&A...696A..81B}. The cosmic component can be further decomposed into a mean, which traces the diffuse baryon density~\citep{2020Natur.581..391M}, and a fluctuation sourced by collapsed structures~\citep{2014ApJ...780L..33M, 2024ApJ...965...57B}. The statistics of these DM perturbations are governed by the baryon power spectrum, and are therefore directly sensitive to the strength of feedback~\citep{2014ApJ...780L..33M, 2025ApJ...983...46M, 2025ApJ...989...81S}. Crucially, FRBs provide a direct probe of the total ionized gas content, regardless of its temperature or density phase. In this respect, FRBs are complementary to probes such as the kinematic Sunyaev-Zel'dovich (kSZ) effect and patchy screening~\citep{2025PhRvD.112j3532H}, which likewise trace ionized gas that contributes negligibly to X-ray emission or the thermal Sunyaev Zel'dovich (tSZ) effect signal.

Analysis of FRB samples have so far primarily exploited the DM-redshift relation (one-point statistic) to measure the baryon density~\citep{2020Natur.581..391M}, and partition baryons between the CGM of halos and the IGM~\citep{2024MNRAS.529..537K, 2025NatAs...9.1226C}. More recently, the sightline-to-sightline variance in DM has been recognized as a sensitive probe of feedback-induced suppression in the matter power spectrum~\citep{2025ApJ...989...81S, 2025arXiv250717742R,2026arXiv260417162S}. These one-point statistic analyses are, however, susceptible to observational selection effects from the FRB luminosity function, density fluctuations-dependent scattering, and instrument detection efficiency that modulate the observed sample~\citep{2019MNRAS.487.5753C, 2022MNRAS.509.4775J, 2022MNRAS.516.4862J, 2026ApJ...999..202S}.

The angular cross-correlation of FRB DMs with large-scale structure (LSS) and baryon tracers sidesteps these limitations. In this framework, FRBs act as \textit{backlights}, analogous to background source galaxies in weak lensing, and the cross-correlation isolates the electron distribution at the location of foreground structures from the DM field of background FRBs. Since the host galaxy contributions, scattering and DM-dependent selection effects are uncorrelated with foreground tracer positions, they do not bias the cross-correlation signal as long as they are well-separated in redshift~\citep{2026ApJ...998..252C}. This robustness to selection effects is a fundamental advantage over one-point statistics, and the feasibility of this approach has been demonstrated by the recent measurements of galaxy$\times$DM~\citep{2025arXiv250608932W, 2025ApJ...993L..27H, 2026arXiv260121336S} and tSZ$\times$DM~\citep{2025arXiv251102155T} correlations.

\begin{figure}
\centering
\includegraphics[width=\columnwidth]{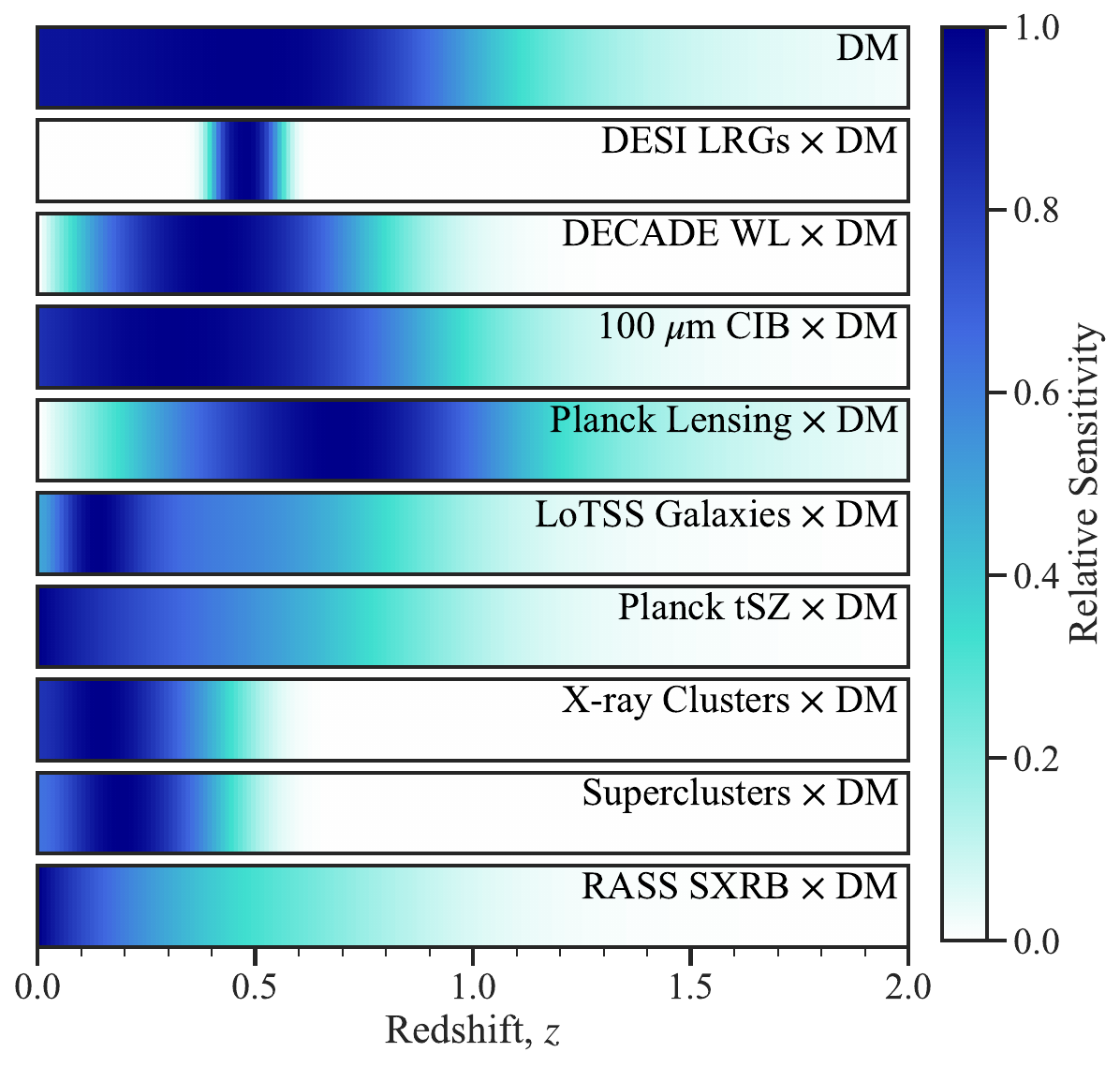}
\caption{Redshift sensitivity of CHIME/FRB DMs~\citep{2026ApJS..283...34C}, when cross-correlated with LSS and baryon tracers. Each panel shows the normalized DM perturbation weighting function convolved with the redshift kernel of various probes, including DESI LRGs~\citep{2023JCAP...11..097Z}, DECADE tangential shear~\citep{2025arXiv250903798G, 2025OJAp....846161A}, WISE-reconstructed 100~$\mu$m CIB~\citep{2023ApJ...958..118C, 2025ApJ...992...65C}, Planck CMB lensing convergence~\citep{2020A&A...641A...8P}, LoTSS star-forming galaxies and AGNs~\citep{2026A&A...707A.198S}, Planck tSZ Compton-$y$ parameter~\citep{2020A&A...643A..42P}, RASS~\citep{2023MNRAS.526.3757K} and eRASS1~\citep{2024A&A...685A.106B} X-ray galaxy groups, clusters and superclusters, and RASS soft X-ray background~\citep[SXRB;][]{1997ApJ...485..125S, 2024PhRvL.133e1001F}. This set of cross-correlations probes the distribution of matter and baryons at redshifts $z \lesssim 1.5$, with different tracers preferentially weighting distinct redshift ranges.}
\label{fig:redshift_sensitivity}
\end{figure}

In this work, we exploit this approach across the broadest suite of FRB DM cross-correlations attempted to date, using 3,455 unique FRB sources from the Canadian Hydrogen Intensity Mapping Experiment Fast Radio Burst project (CHIME/FRB) Catalog 2, with a median localization uncertainty of $\sim 15$~arcmin~\citep{2026ApJS..283...34C}. We cross-correlate with ten tracers that together probe complementary aspects of the baryon-LSS connection across a wide range of physical environments and redshifts. The redshift sensitivity of CHIME/FRB DM field, when cross-correlated with these tracers, is illustrated in Figure~\ref{fig:redshift_sensitivity}, which shows the normalized DM perturbation weighting function multiplied by the radial kernel of each probe (see Section~\ref{sec:analysis_framework} for detailed discussion of the theoretical framework).

For LSS, we use the first photometric tomography bin of Dark Energy Spectroscopic Instrument (DESI) Luminous Red Galaxies~\citep[LRGs;][]{2023AJ....165...58Z}, which provide a well-characterized tracer of the matter field at $z \sim 0.4-0.55$; Dark Energy Camera All Data Everywhere (DECADE) weak lensing~\citep{2025arXiv250903798G, 2025OJAp....846159A}, which probes the line-of-sight integral of the matter distribution out to $z \sim 1.5$; 100~$\mu$m cosmic infrared background (CIB), which probes dusty star-forming galaxies at $z \lesssim 3$~\citep{2023ApJ...958..118C, 2025ApJ...992...65C}; Planck cosmic microwave background (CMB) lensing convergence~\citep{2020A&A...641A...8P}, which extends sensitivity out to the surface of last scattering; and LOw-Frequency ARray Two-metre Sky Survey (LoTSS) continuum flux density maps and source catalogs~\citep{2026A&A...707A.198S}, which trace a combination of AGNs and star-forming galaxies at $z \lesssim 5$. For thermal and ionized gas, we use the Planck tSZ Compton-$y$ map, which traces the pressure-weighted electron distribution~\citep{2016A&A...594A..22P}; ROSAT all-sky survey (RASS) and eROSITA all-sky survey (eRASS1) X-ray galaxy groups and clusters~\citep{2023MNRAS.526.3757K, 2024A&A...685A.106B}, probing the hot, dense intracluster medium (ICM); eRASS1 superclusters~\citep{2024A&A...683A.130L}, extending bandwidth to the largest gravitationally bound structures; and the RASS soft X-ray background~\citep{1997ApJ...485..125S, 2024PhRvL.133e1001F}, capturing the integrated emission from diffuse hot gas. Together, these tracers span the baryon distribution from densest galaxy clusters to diffuse IGM.

The remainder of this paper is organized as follows. We begin by describing the datasets used in our analysis in Section~\ref{sec:data}, followed by our analysis framework in Section~\ref{sec:analysis_framework}, including theoretical models, cross-correlation estimators, and covariance measurement procedures. We present correlation functions and measurement statistics in Section~\ref{sec:measured_correlation_functions}. We discuss the physical implications of our measurements in Section~\ref{sec:discussion} and conclude in Section~\ref{sec:conclusion}. Throughout this work, we use the \citet{2020A&A...641A...6P} TT$+$TE$+$EE$+$lowE cosmology.

\section{Data Description} \label{sec:data}

In this section, we describe the data products used in our measurements, including the FRB DMs from CHIME/FRB Catalog 2 (Section~\ref{subsec:DM_from_CHIME_cat2}), and the tracers of LSS and baryonic matter (Section~\ref{subsec:LSS_baryon_tracers}). The gallery of all tracers used in this work is shown in Figure~\ref{fig:maps}. 

\begin{figure*}[ht!]
\centering
\includegraphics[width=0.325\textwidth]{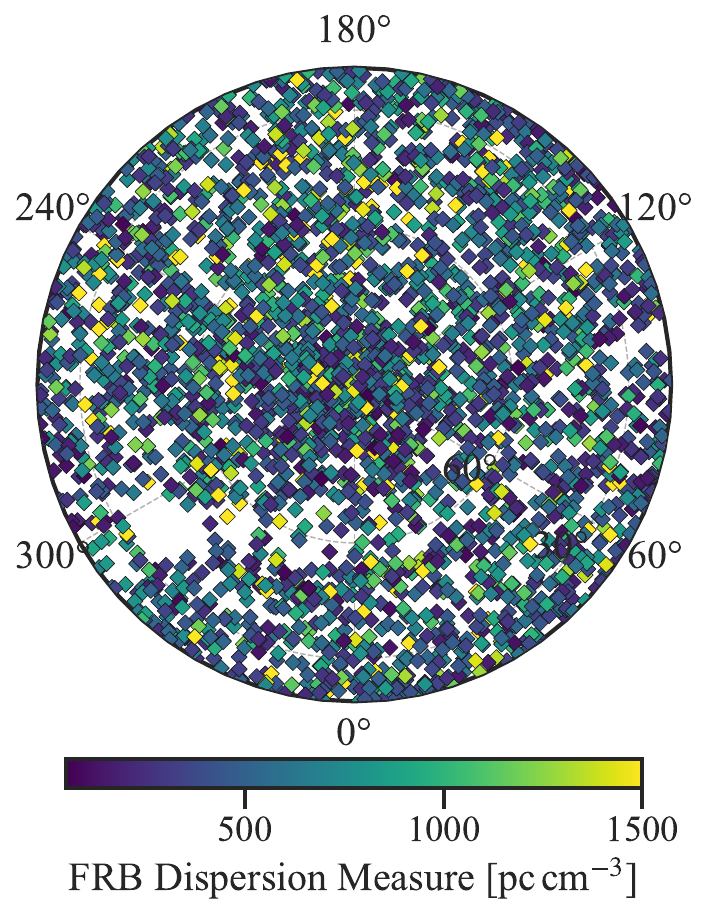}
\includegraphics[width=0.325\textwidth]{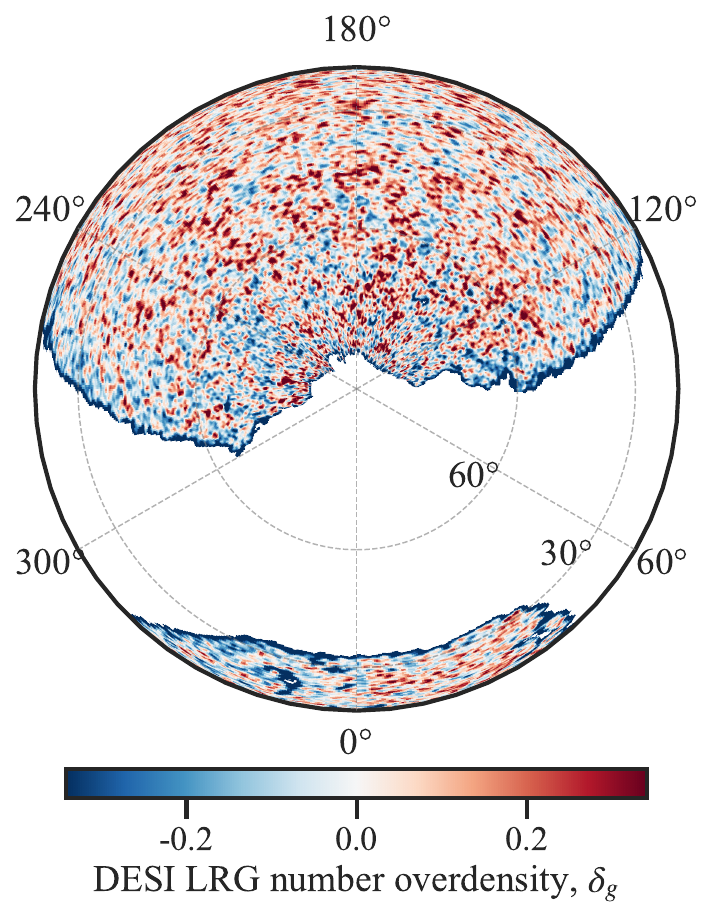}
\includegraphics[width=0.325\textwidth]{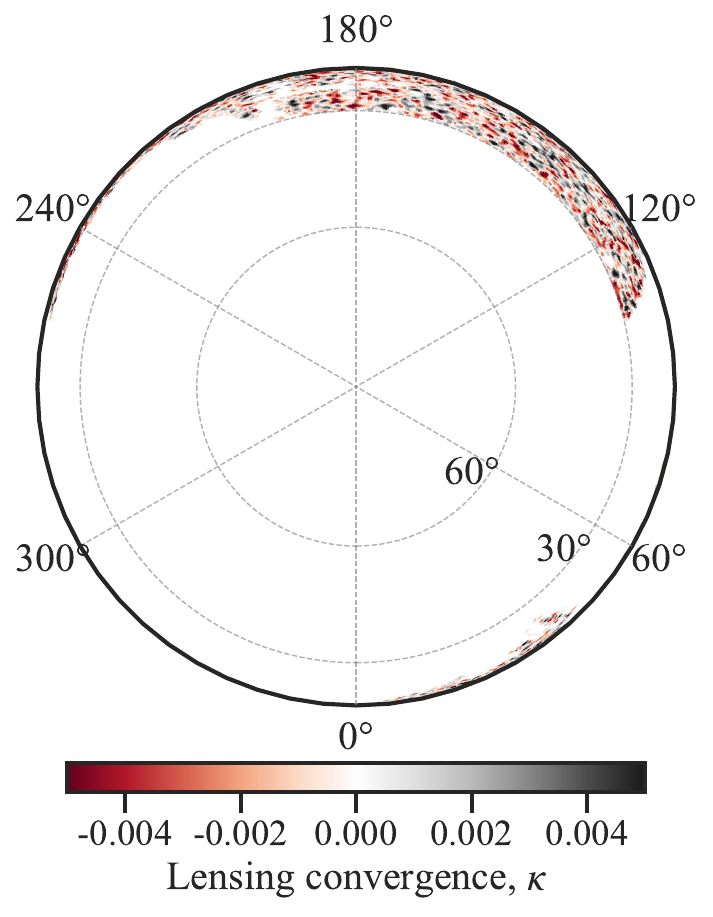}
\includegraphics[width=0.325\textwidth]{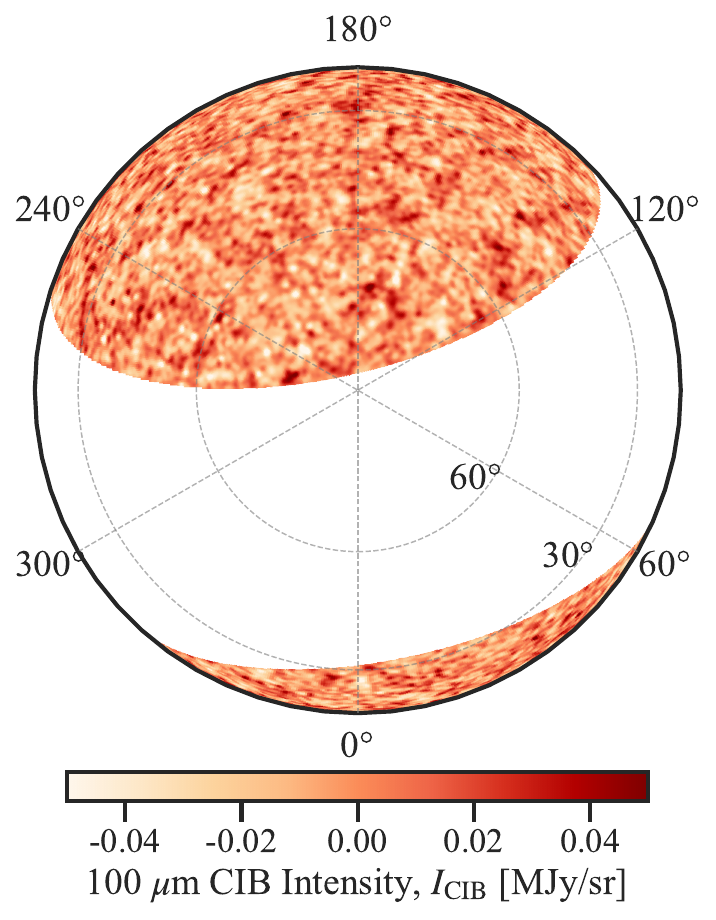}
\includegraphics[width=0.325\textwidth]{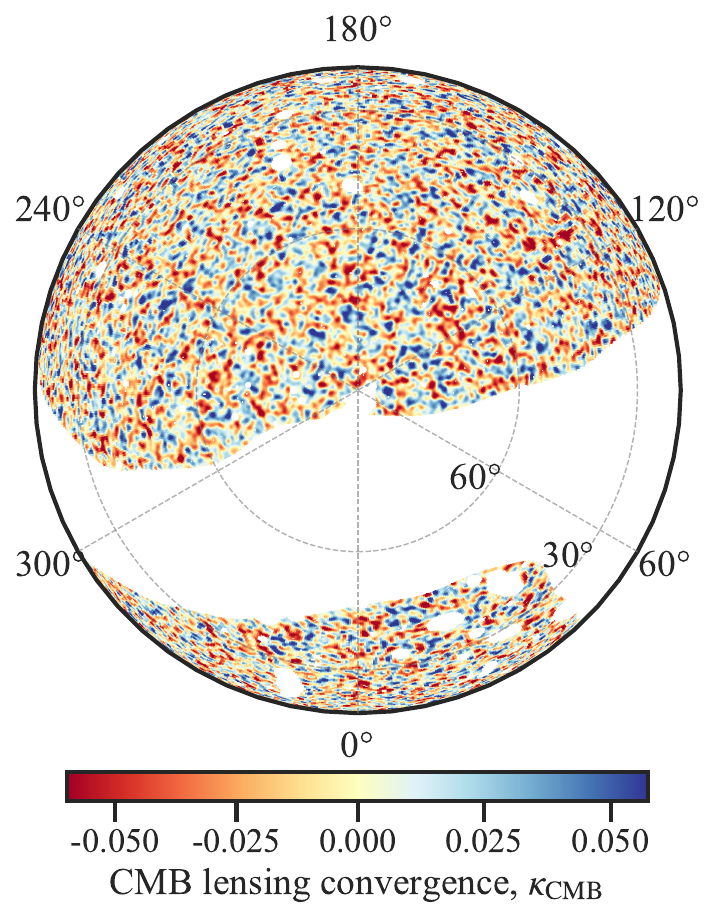}
\includegraphics[width=0.325\textwidth]{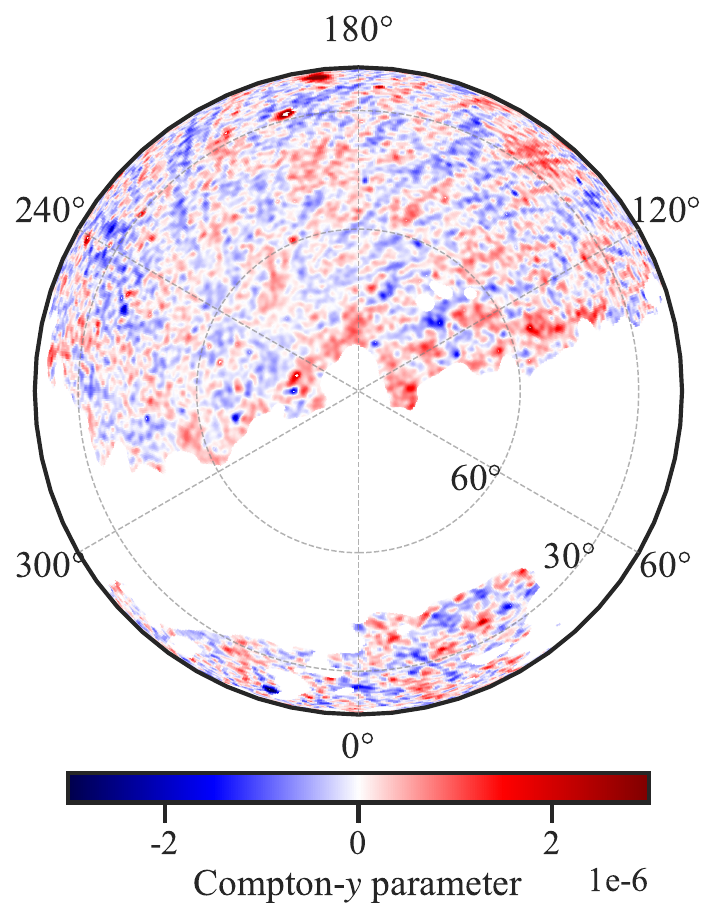}
\caption{Sky maps in an orthographic projection centered on the north celestial pole, showing the CHIME/FRB sample~\citep{2026ApJS..283...34C}, alongside the ten LSS and baryons tracers cross-correlated in this work. Dashed grid lines mark declinations of 30$^\circ$ and 60$^\circ$, and right ascension intervals of 60$^\circ$. Unobserved or masked regions are left blank. All maps are smoothed with a $1^\circ$ FWHM Gaussian for display. \textit{Upper panels}: CHIME/FRB catalog, with each FRB colored by its NE2001-subtracted DM; the DESI LRG number overdensity ($\delta_g$)~\citep{2023JCAP...11..097Z}; DECADE weak lensing convergence ($\kappa$) map~\citep{2025arXiv250903798G, 2025OJAp....846161A}. \textit{Lower panels}: WISE-reconstructed 100~$\mu$m CIB~\citep{2023ApJ...958..118C} intensity ($I_\mathrm{CIB}$); Planck PR3 tSZ-deprojected CMB lensing convergence ($\kappa_\mathrm{CMB}$)~\citep{2020A&A...641A...8P}; Planck Compton-$y$ map~\citep{2020A&A...643A..42P}.}
\label{fig:maps}
\end{figure*}

\begin{figure*}
\addtocounter{figure}{-1}
\centering
\includegraphics[width=0.325\textwidth]{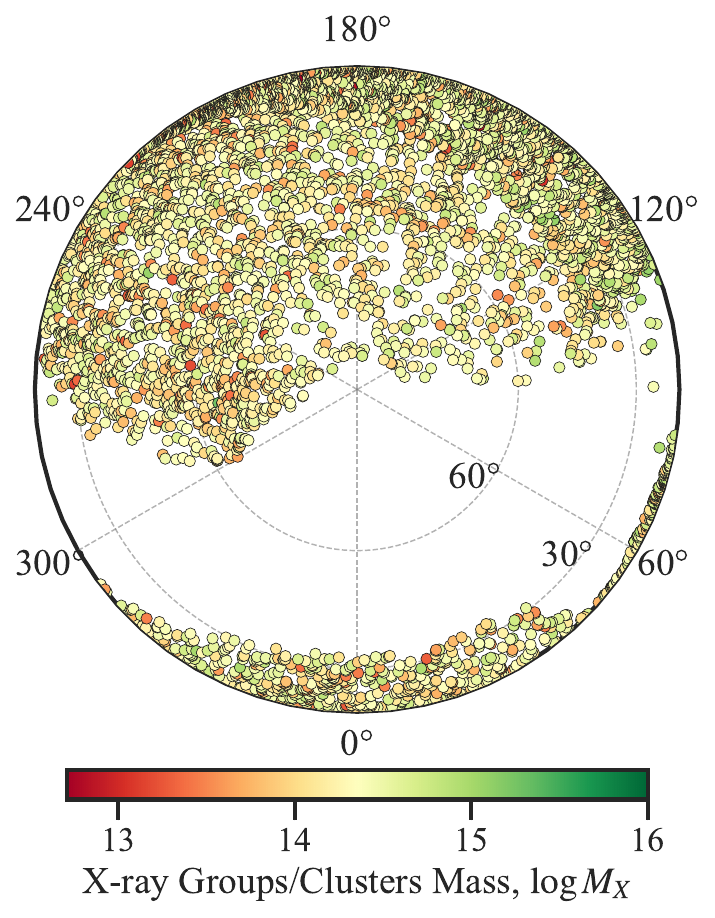}
\includegraphics[width=0.325\textwidth]{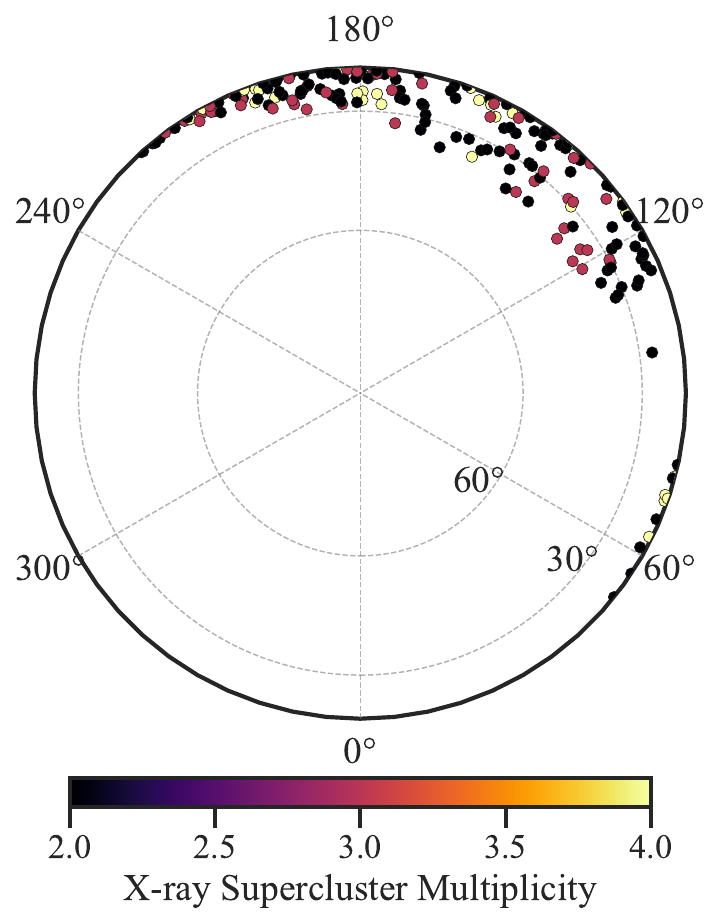}
\includegraphics[width=0.325\textwidth]{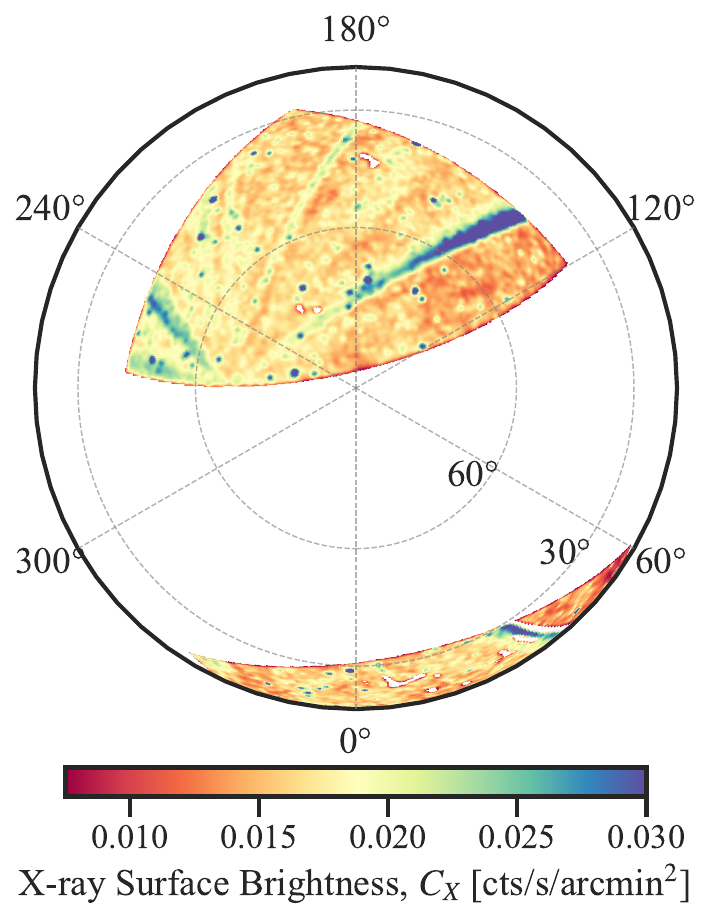}
\includegraphics[width=0.325\textwidth]{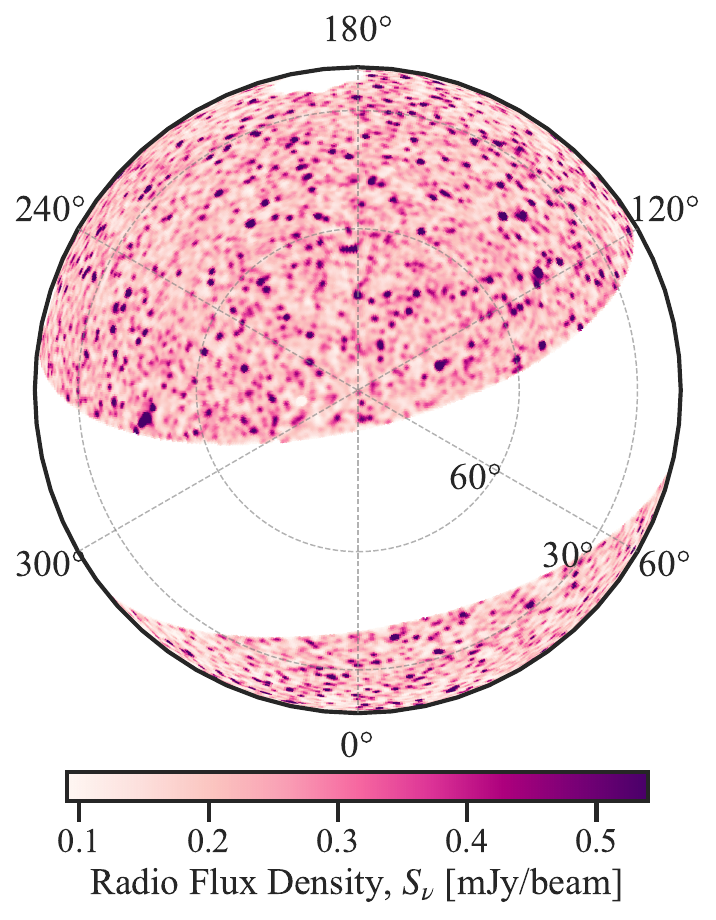}
\includegraphics[width=0.325\textwidth]{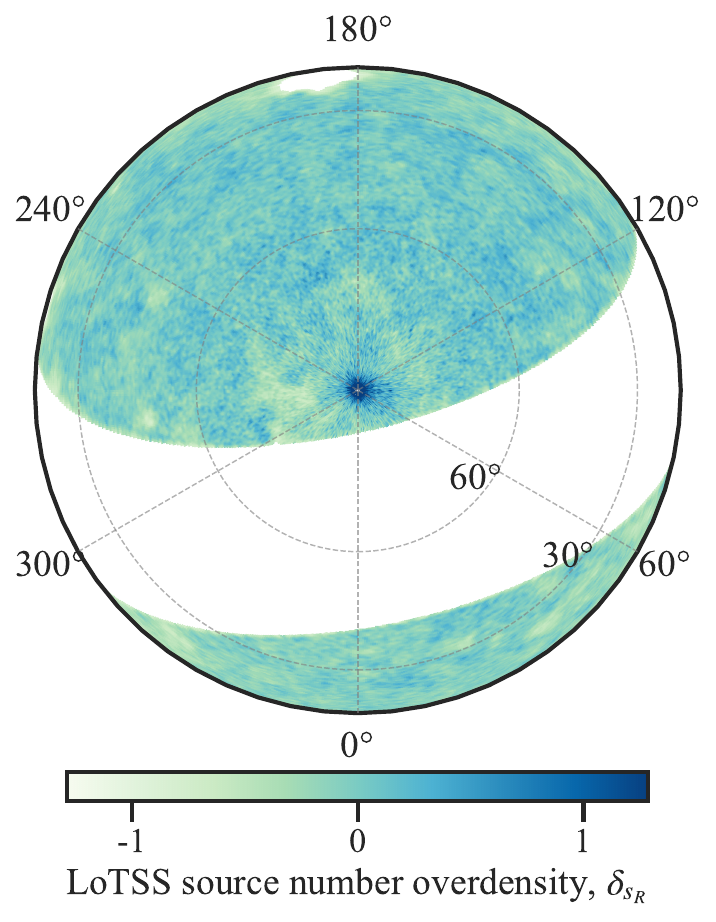}
\caption{\textit{(Cont.)} \textit{Upper panels}: RASS~\citep{2023MNRAS.526.3757K} and eRASS1 ~\citep{2024A&A...685A.106B} X-ray galaxy groups and clusters, colored by mass; eRASS1 superclusters at $z <0.5$, colored by multiplicity~\citep{2024A&A...683A.130L}; ROSAT soft X-ray background~\citep{1997ApJ...485..125S, 2024PhRvL.133e1001F}. \textit{Bottom panels}: LoTSS radio sky flux density~\citep{2026A&A...707A.198S}; LoTSS radio source number overdensity~\citep{2026A&A...707A.198S}.}
\end{figure*}

\subsection{Dispersion Measures from CHIME/FRB Catalog 2} \label{subsec:DM_from_CHIME_cat2}

We use DM measurements from the second catalog of the CHIME/FRB project~\citep{2026ApJS..283...34C}. CHIME functions as a transit telescope, providing near-uniform daily sky coverage over the declination range accessible from its latitude, with exposure increasing towards the north celestial pole. The CHIME/FRB Catalog 2 contains 5,045 FRBs, originating from 3,558 unique sources, making it the largest FRB sample from a single instrument to date.

We apply the following selection cuts to construct the FRB sample used in this work. We retain events with well-measured sky positions and refined DM estimates from the \textsc{fitburst} spectro-temporal modeling pipeline~\citep{2024ApJS..271...49F}, removing 103 FRBs with missing measurements of either quantity. We restrict the sample to single burst per unique source identified by its Transient Name Server designation. Where multiple sub-bursts from the same event are present, we retain only the primary burst. After these cuts, the sample comprises 3,455 sightlines. The localization uncertainty of this sample spans $0.2-32$~arcmin with a median of $\sim 15$~arcmin. We therefore limit our analysis to $\gtrsim 10$~arcmin angular scales.

For cross-correlations with extragalactic tracers, we use the NE2001-subtracted estimates of extragalactic DM, DM$_\mathrm{ex} = \mathrm{DM}_\mathrm{FITB} - \mathrm{DM}_\mathrm{NE2001}$, where $\mathrm{DM}_\mathrm{FITB}$ is the \textsc{fitburst} DM estimate and $\mathrm{DM}_\mathrm{NE2001}$ is the Galactic DM predicted by the NE2001 electron density model~\citep{2002astro.ph..7156C} along each sightline. The sky distribution of the final FRB sample, colored by extragalactic DM, is shown in Figure~\ref{fig:maps}. The sample covers the northern sky accessible to CHIME, with the concentration of events towards the north celestial pole reflecting the telescope's increased exposure at high declinations as a transit instrument. The under-density of FRBs along the Galactic plane arises from reduced observational sensitivity along sightlines with significant scattering, caused by regions of large electron density~\citep{2026ApJ...997L...5P}. 

\subsection{Large-Scale Structure and Baryon Tracers} \label{subsec:LSS_baryon_tracers}

In this section, we summarize the diverse set of LSS and baryon tracers used in this work, where each probe preferentially weights distinct redshift ranges and physical regimes of the baryon distribution (see Figure~\ref{fig:redshift_sensitivity}).

\subsubsection{DESI Luminous Red Galaxies} \label{subsubsec:DESI_LRGs}

We use the photometric LRG catalog presented in \citet{2023JCAP...11..097Z}, which is drawn from the DESI Legacy Imaging Surveys Data Release (DR) 9~\citep{2021AAS...23723503S}. The catalog covers approximately 19,700~deg$^2$ of extragalactic sky. The selection criteria follow those of the DESI Main Survey LRG sample~\citep{2023AJ....165...58Z}, employing cuts to reject stellar sources, exclude low-redshift galaxies, and limit luminosity that produces a comoving number density of $5 \times 10^{-4}h^3\,$\,Mpc$^{-3}$ at redshifts $ 0.4 < z < 0.8$~\citep{2023AJ....165...58Z}. The random catalog is constructed following the same pipeline. Galaxies are divided into four tomography bins, out of which we only use the lowest-redshift bin which spans $0.4 \lesssim z \lesssim 0.55$ to maximize the probability that galaxies lie in front of the FRBs, given their DM distribution. While an analogous galaxy$\times$DM correlation with lower-redshift bright galaxy sample (BGS) has already been conducted by \citet{2025arXiv250608932W}, our use of the LRG sample extends the set of such measurements, and is particularly compelling given the complementary kSZ measurements from Atacama Cosmology Telescope (ACT) for this galaxy sample~\citep{2025PhRvD.112l3507H, 2025PhRvD.112h3509H}. The redshift distribution~\citep{2023AJ....165...58Z} and angular clustering~\citep{2023AJ....165...58Z, 2024MNRAS.530..947Y} of DESI LRGs have been characterized using the DESI Year 1 spectroscopic redshifts.

\subsubsection{DECADE Weak Gravitational Lensing} \label{subsubsec:DECADE_WL}

We use the galaxy shape catalog from Dark Energy Camera All Data Everywhere (DECADE) cosmic shear project~\citep{2025OJAp....846158A}, which provides weak lensing measurements across $\sim$13,000~deg$^2$. The catalog contains $\sim$270~million galaxies divided into four tomography bins with mean redshifts $\langle z \rangle \in \{0.33, 0.50, 0.71, 0.91\}$~\citep{2025OJAp....846159A}. Full details of the catalog construction, shear calibration, and redshift estimation are provided in \citet{2025OJAp....846159A, 2025OJAp....846158A, 2025OJAp....846160A}. The source galaxy catalog provides per-galaxy ellipticity components $(g_1, g_2)$ and inverse-variance weights from meta-calibration. In this work, we restrict our analysis to the fourth tomography bin, as its broader lensing kernel provides better overlap with the DM perturbation weighting function.

\subsubsection{100~$\mu$m Cosmic Infrared Background} \label{subsubsec:WISE_CIB}

We use the 100~$\mu$m CIB intensity map from \citet{2023ApJ...958..118C}, which isolates the extragalactic CIB component in the \citet{1998ApJ...500..525S} dust map through a tomographic field-level reconstruction using $\sim 600$~million WISE galaxies~\citep{2010AJ....140.1868W, 2016ApJS..225....5B, 2021ApJS..253....8M} as LSS templates, yielding a foreground-free CIB reconstruction spanning $z \sim 0-3$, with point sources masked. The CIB consists of integrated infrared emission from unresolved dusty star-forming galaxies, with most emission originating from $z \lesssim 2$, making it a biased tracer of LSS. We mask regions with Galactic latitude $|b| < 30^\circ$ to minimize residual Galactic dust contamination, at the cost of a reduced sky fraction. This conservative choice ensures that any measured DM-CIB correlation reflects genuine LSS rather than Galactic foreground leakage.

\subsubsection{Planck CMB Lensing Convergence} \label{subsubsec:Planck_CMB_lensing}

We use CMB lensing convergence maps from the Planck 2018 data release~\citep{2020A&A...641A...8P}, constructed with temperature-only (TT), polarization-only (PP), and minimum variance (MV) quadratic estimators applied to foreground-cleaned SMICA CMB maps. Our primary lensing map uses the tSZ-deprojected TT-only reconstruction, which is built by input T maps formed using frequency weights that null the tSZ component. Using the standard (non-tSZ-deprojected) lensing reconstruction can modify the inferred $\kappa_\mathrm{CMB}$$\times$DM signal through two related effects. First, residual foreground contamination alters the reconstruction noise and response, typically redistributing power between intermediate-to-large ($\ell \lesssim 2000$) and small ($\ell \gtrsim 2000$) angular scales, thereby impacting the measured cross-spectrum. Second, the presence of residual tSZ emission, spatially correlated with high-density regions traced by $\kappa_\mathrm{CMB}$, introduces an additional contribution to the reconstructed lensing field through mode-coupling terms. This contribution does not have a definite sign a priori and can therefore bias the inferred $\kappa_\mathrm{CMB}$$\times$DM amplitude either upward or downward, depending on scale and weighting. A similar consideration applies to residual CIB contamination. The tSZ deprojected reconstruction removes this contamination at the cost of slightly increased noise. While this effect is expected to be subdominant, entering only through a bispectrum term that is likely negligible, we nevertheless test it explicitly.

We construct the lensing convergence map from the \citet{2020A&A...641A...8P} lensing potential harmonic coefficients, smoothed with a 7~arcmin FWHM Gaussian beam, retaining only modes in the range $40 \leq \ell \leq 2048$. The lower cut at $\ell = 40$ removes the largest angular scales where the mean-field subtraction in the quadratic estimator is less reliable, and the upper cut at $\ell = 2048$ avoids the smallest scales where reconstruction noise dominates and residual foreground contamination is significant. The smoothed, band-limited harmonic coefficients are then projected to a \textsc{HEALPix} map with $N_\mathrm{side}=2048$. We mask the Galactic plane, bright point sources, and resolved SZ clusters.

\subsubsection{Planck Thermal Sunyaev-Zel’dovich Effect} \label{subsubsec:Planck_tSZ}

We use the full-mission Planck NILC Compton-$y$ map from the 2015 data release~\citep{2016A&A...594A..22P}, reconstructed at $N_\mathrm{side}=2048$ with a 10~arcmin FWHM beam. The NILC algorithm combines the Planck HFI channel maps from 100 to 857~GHz in needlet space, imposing unit gain on the tSZ spectral signature while simultaneously nulling the CMB contribution~\citep{2011MNRAS.418..467R}. The dominant sources of contamination at small angular scales include residual extragalactic point sources (radio or infrared), which are masked using a union of individual frequency-channel point source masks. The processed NILC maps are known to be dominated by the tSZ signal in multipole range $20 < \ell < 600$~\citep{2016A&A...594A..22P}. The Planck $y$-map reconstructions exhibit a negative $\mathrm{d}y/\mathrm{d}z$ at $z \sim 1$ and an overshoot at higher redshifts, consistent with residual CIB contamination: the ILC procedure, which nulls the mean CIB, induces an artificial anti-correlation at intermediate redshift~\citep{2020ApJ...902...56C}. Since our tSZ$\times$DM signal is dominated by $z \lesssim 1$ (see Figure~\ref{fig:redshift_sensitivity}), these high-redshift artifacts have limited impact here, but will become important for higher-redshift FRB samples in future.

\subsubsection{RASS and eRASS1 X-ray Galaxy Clusters, Superclusters and Soft X-ray Background} \label{subsubsec:RASS_clusters_SXRB}

We use three X-ray selected galaxy cluster and supercluster catalogs, together with the diffuse SXRB map. For the primary cluster sample, we use \citet{2023MNRAS.526.3757K} catalog, constructed by applying a multi-component matched filter cluster confirmation algorithm to the second RASS source catalog~\citep{2016A&A...588A.103B} over 25,000~deg$^2$ covered by DESI DR10. The algorithm identifies optical counterparts of X-ray sources by searching for red-sequence galaxy over-densities, and measuring photometric redshift, richness and contamination probability. The resulting 90\% purity catalog contains 8,449 X-ray-selected galaxy clusters, with a redshift distribution peaking at $z\sim 0.2$. In this work, we restrict to 8,034 clusters at redshifts $z < 0.5$, effectively making them a transmissive disperser for our FRB sample. We further restrict to 7,284 clusters at Galactic latitude $|b| > 25^\circ$, out of which 4,898 clusters lie within a 10$^\circ$ radius of our FRB sample.

We supplement the RASS sample with galaxy clusters from eRASS1~\citep{2024A&A...685A.106B}. The eRASS1 identifies 12,247 extended galaxy clusters and groups over 13,116~deg$^2$, optically confirmed using the eROMaPPer red-sequence cluster finder applied to DESI DR9 and DR10, with contamination estimated via a mixture model exploiting X-ray count rates and richness. We apply several quality cuts to the primary eRASS1 sample to construct a higher-purity subsample~\citep{2024A&A...683A.130L}. We restrict to 6,302 clusters with contamination probability $P_\mathrm{cont} < 0.3$, redshift uncertainty $\delta z/(1+z) < 0.01$, at least 5 X-ray photons detected within $R_{500}$, photometric redshifts within the limiting redshift range of the instrument at their sky position, redshifts in the range $0.005 < z < 0.5$, and Galactic latitude $|b|>25^\circ$. From this subsample, only 1,931 sources are located within a 10$^\circ$ radius of FRB sources. We treat the two cluster catalogs as independent disperser samples in separate cross-correlation measurements.

In addition to individual clusters, we also consider 1,338 eRASS1 superclusters~\citep{2024A&A...683A.130L}, identified by applying Friends-of-Friends algorithm to the higher-purity subsample of primary eRASS1 catalog. These include 818 cluster pairs and 520 rich systems ($\geq 3$ members) with mass in the range $\sim 6 \times 10^{13} - 2 \times 10^{16}~M_\odot$, and projected length upto $\sim 130$~Mpc. For our analysis, we restrict to 310 systems at $z < 0.5$ and Galactic latitude $|b|>25^\circ$. 

As a complementary probe of diffuse hot gas in the outskirts of galaxy groups and clusters, and the warm-hot gas in filaments, we use the SXRB map from RASS~\citep{1997ApJ...485..125S, 2024PhRvL.133e1001F}. The count rate (cts~s$^{-1}$arcmin$^{-2}$) maps are constructed in 0.5-2.0~keV band by \citet{2024PhRvL.133e1001F} in \textsc{HEALPix}-projection at $N_\mathrm{side} = 1024$ resolution. We mask pixels with exposure time less than 100~seconds to avoid regions of poor sensitivity. To suppress Galactic foreground contamination, we mask pixels with $|b|<30^\circ$, and known extended features at $l = 70-240^\circ$ from Loop I/North Polar Spur.

\subsubsection{LoTSS Radio Sky and Radio Sources} \label{subsubsec:LoTSS_radio_sky_sources}

We use data products from LoTSS DR3~\citep{2026A&A...707A.198S}, which imaged 88\% of the northern sky at 120-168~MHz, achieving a median sensitivity of 92~$\mu$Jy/beam at 6$\arcsec$ angular resolution. LoTSS-DR3 catalogued 13~million radio sources extracted from mosaicked continuum images, tracing a combination of AGNs and star-forming galaxies. Cross-match of LoTSS Deep Fields~\citep{2021A&A...648A...6S} with galaxy surveys has shown that the redshift distribution of sources with flux density $S_\nu \geq 1.5$~mJy peaks at $z \sim 0.8$ and extends to $z \sim 5$~\citep{2024MNRAS.527.6540H, 2024A&A...681A.105N}.

To cross-correlate with large-scale radio emission from AGNs and star-forming galaxies, we use the 20$\arcsec$ resolution continuum mosaics, reprojected into a \textsc{HEALPix} map at $N_\mathrm{side} = 512$ resolution. We also correlate DMs with individual radio sources from the LoTSS-DR3 catalog weighted by their observed flux density~\citep{2026A&A...707A.198S}. Following the selection criteria established for LoTSS cosmological analyses~\citep{2024MNRAS.527.6540H, 2024A&A...681A.105N}, we restrict to sources with $S_\nu > 1.5$~mJy and peak signal-to-noise ratio $S_\mathrm{peak}/S_\mathrm{rms} > 7.5$. At fainter flux densities, residual field-to-field systematics in the flux density scale and incompleteness corrections render the catalog unsuitable for LSS studies~\citep{2024MNRAS.527.6540H}. Furthermore, we suppress the Galactic foreground contamination by masking $|b| < 20^\circ$.

\section{Analysis Framework} \label{sec:analysis_framework}

We describe the analysis framework used to measure and interpret the cross-correlations of CHIME/FRB DMs with LSS and baryon tracers described in Section~\ref{sec:data}. We first outline the theoretical models used to predict the expected cross-correlation signal in Section~\ref{subsec:theory_models}, followed by a description of the cross-correlation estimators in Section~\ref{subsec:cross_correlation_estimators}, covariance estimation procedure in Section~\ref{subsec:jackknife_covariance_estimation}, and finally the signal-to-noise metric used to quantify the measurement significance in Section~\ref{subsec:SNR_estimation}.

\subsection{Theoretical Models} \label{subsec:theory_models}

The theoretical model for angular correlation function $\xi^{AB}(\theta)$ between two projected fields $A(\hat{x})$ and $B(\hat{x})$ is related to the angular cross-power spectrum $C^{AB}(\ell)$ via a Hankel transform:
\begin{equation}
\xi^{AB}(\theta) = \int\limits_0^\infty \frac{\ell \, \mathrm{d}\ell}{2\pi} \, C^{AB}(\ell) \, J_{n}(\ell \theta),
\label{eqn:theoretical_correlation_function}
\end{equation}
where $J_n$ is the $n$th-order Bessel function of the first kind, with $n=0$ for scalar, and $n=2$ for tangential shear cross-correlations, $\ell$ denotes the multipole, and $\theta$ is the angular separation on sky. Under Limber approximation, the angular cross-power spectrum is
\begin{equation}
C^{AB} (\ell) = \int\limits_0^\infty \frac{\mathrm{d}\chi}{\chi^2} \, W_A(\chi) \, W_B(\chi) \, P_{AB}\Big(k = \frac{\ell}{\chi}, z(\chi)\Big),
\label{eqn:angular_power_spectrum}
\end{equation}
where $\chi$ is the comoving distance, $P_{AB}(k,z)$ is the three-dimensional cross-power spectrum of fields $A$ and $B$, and $W_A(\chi)$, $W_B(\chi)$ are their respective line-of-sight weighting kernels. The Limber approximation is valid for sufficiently large $\ell$, where the spherical Bessel functions are
peaked around $k\chi = \ell$. Within the halo model framework, the cross-power spectrum of fields $A$ and $B$ decomposes into a one-halo (1h) and a two-halo (2h) term,
\begin{equation}
P_{AB}(k,z) = P_{AB}^\mathrm{1h}(k,z) + P_{AB}^\mathrm{2h}(k,z),
\label{eqn:halo_model_power_spectrum}
\end{equation}
where the one-halo term captures contributions from clustering within the same halo,
\begin{equation}
P_{AB}^\mathrm{1h}(k,z) = \int_0^\infty u_A(M,k)\, u_B(M,k)\, n(M,z)\, \mathrm{d}M,
\label{eqn:1halo_term}
\end{equation}
and the two-halo term encodes the large-scale clustering of distinct halos,
\begin{equation}
\begin{aligned}
P&_{AB}^\mathrm{2h}(k,z) = \left[\int_0^\infty b(M,z)\, u_A(M,k)\, n(M,z)\, \mathrm{d}M\right] \\
& \left[\int_0^\infty b(M,z)\, u_B(M,k)\, n(M,z)\, \mathrm{d}M\right] P_{AB}^\mathrm{lin}(k,z).
\end{aligned}
\label{eqn:2halo_term}
\end{equation}
Here, $n(M,z)$ is the halo mass function~\citep{2008ApJ...688..709T}, $b(M,z)$ is the large-scale halo bias~\citep{2010ApJ...724..878T}, and $u_v(M,k)$ is the spherical Fourier transform of the halo profile for field $v$. We use halo profiles from the analytical model \textsc{BCEmu}~\citep{2019JCAP...03..020S, 2021JCAP...12..046G}, as implemented in \textsc{BaryonForge}\footnote{\url{https://github.com/DhayaaAnbajagane/BaryonForge}}~\citep{2024OJAp....7E.108A}, and compute power spectra using \textsc{pyccl}\footnote{\url{https://github.com/LSSTDESC/CCL}}.  Below, we summarize the weighting functions and halo profiles for each probe.

\subsubsection{FRB Dispersion Measures}

The DM perturbation field at angular position $\hat{x}$, sourced by fluctuations in the electron density $\delta_\mathrm{e}$, is
\begin{equation}
\mathcal{D}(\hat{x}) = \int_0^{\chi_\mathrm{H}} \mathrm{d}\chi\, W_\mathcal{D}(\chi)\, \delta_\mathrm{e}(\hat{x}, \chi),
\label{eqn:DM_perturbation_field}
\end{equation}
where $\chi_\mathrm{H}$ is the comoving distance to the horizon, and $W_\mathcal{D}(\chi)$ is the DM perturbation weighting function~\citep{2023arXiv230909766R, 2025ApJ...989...81S, 2026ApJ...998..109S},
\begin{equation}
W_{\mathcal{D}}(\chi) = W_\mathrm{DM}(\chi) 
\int\limits_\chi^{\chi_\mathrm{H}} \mathrm{d}\chi^\prime \, \frac{n_\mathrm{f}(z(\chi^\prime))}{\bar{n}_\mathrm{f}} \frac{\mathrm{d}z}{\mathrm{d}\chi^\prime},
\label{eqn:DM_perturbation_wt_fxn}
\end{equation}
with $n_\mathrm{f}(z)$ and $\bar{n}_\mathrm{f}$ the redshift distribution and mean angular number density of FRBs, respectively, and $W_\mathrm{DM}(\chi)$ encodes the line-of-sight response of DM:
\begin{equation}
W_\mathrm{DM}(\chi) = \frac{3c \chi_\mathrm{e} \Omega_\mathrm{b} H_0}{8 \pi G m_\mathrm{p}} \frac{f_\mathrm{diffuse}(z) (1+z)}{\sqrt{\Omega_\mathrm{m}(1+z)^3 + \Omega_{\Lambda}}} \frac{\mathrm{d}z}{\mathrm{d}\chi}.
\label{eqn:DM_response_wt_fxn}
\end{equation}
Here, $f_\mathrm{diffuse}(z)$ is the fraction of baryons not locked in stars and atomic/molecular hydrogen gas~\citep{2025A&A...695A.163B, 2020ARA&A..58..363P}, and the free electron fraction $\chi_\mathrm{e} = Y_\mathrm{H} + Y_\mathrm{He}/2 \approx 1 - Y_\mathrm{He}/2$ depends on the primordial helium abundance~\citep{2020A&A...641A...6P}. Approximating electron bias $b_\mathrm{e} \approx 1$ (demonstrated to be accurate at $\lesssim$1\%-level upto scales $k \sim 5~h$\,Mpc$^{-1}$~\citep{2025arXiv250919514L}), the electron power spectrum $P_\mathrm{ee}(k,z)$ is equal to the gas power spectrum $P_\mathrm{gas}(k,z)$. Therefore, for cross-power spectrum calculations with electron over-density, the radial gas profile of a halo of mass $M$ is written as
\begin{equation}
\rho_\mathrm{gas}(r|M) = \dfrac{\rho_\mathrm{gas, 0}}{\left[ 1+\left( \dfrac{r}{\theta_\mathrm{co} r_\mathrm{200c}} \right)^\beta \right] \left[ 1+\left( \dfrac{r}{\theta_\mathrm{ej} r_\mathrm{200c}} \right)^\gamma \right]^{\frac{\delta-\beta}{\gamma}}},
\label{eqn:gas_profile}
\end{equation}
where $\rho_\mathrm{gas, 0}$ is a normalization constant. This profile includes a cored inner region (with fixed core scale $\theta_\mathrm{co} = 0.1$), truncated at the gas ejection radius $\theta_\mathrm{ej}r_\mathrm{200c}$. The halo mass dependent slope $\beta$ is
\begin{equation}
\beta = \dfrac{3(M/M_c)^{\mu_\beta}}{1+(M/M_c)^{\mu_\beta}},
\label{eqn:gas_profile_slope}
\end{equation}
where $M_c$ is the characteristic halo mass below which the gas profile becomes shallower than the NFW form and $\mu_\beta$ controls the mass scaling of the slope, accounting for more efficient gas retention in massive clusters relative to groups. The parameters $\gamma$ and $\delta$ govern the slope of the profile beyond the truncation radius. The halo gas mass fraction is set by
\begin{equation}
f_\mathrm{gas}(M) = \frac{\Omega_\mathrm{b}}{\Omega_\mathrm{m}} - f_\mathrm{star}(M),
\label{eqn:f_gas}
\end{equation}
where the stellar mass fraction of a halo is given by 
\begin{equation}
f_\mathrm{star}(M) = 2A \left[\left(\frac{M}{M_1}\right)^{\tau} + \left(\frac{M}{M_1}\right)^{\eta}\right]^{-1},
\label{eqn:f_star}
\end{equation}
with normalization $A=0.055/2$, pivot mass $M_1=2.5\times10^{11}\,M_\odot h^{-1}$~\citep{2013MNRAS.428.3121M}, and power law indices $\tau=-1.5$~\citep{2019MNRAS.488.3143B}, $\eta$ governing the slope of the stellar-to-halo mass relation. We fix other parameters to the median of posterior inferred by \citet{2026arXiv260417162S} from an independent FRB DM-$z$ analysis: $\log M_c = 12.84$, $\mu_\beta = 0.94$, $\delta = 7.33$, $\theta_\mathrm{ej} = 4.93$, $\eta = 0.22$, $\gamma = 2.79$, and $\eta_\delta = 0.21$ (Equation~\ref{eqn:f_cga}). Since the redshifts of FRBs are unknown due to large localization uncertainties, we assign each FRB an approximate redshift assuming the mean DM-redshift relation with aforementioned parameter values. This is justified because we do not conduct inference with these models; rather, we use them only to obtain an estimate of the expected signal from theory. We note that we do not use these synthetic redshifts for subtracting the mean DM contribution from the observed FRB DMs prior to performing cross-correlations. Instead, we remove an ensemble average DM of the sample, as detailed in Section~\ref{subsec:cross_correlation_estimators}.

\subsubsection{Galaxy and Cluster Number Density}

The projected overdensity of a source population along the line-of-sight is
\begin{equation}
\delta_\mathrm{s}(\hat{x}) = \int_0^{\chi_\mathrm{H}} \mathrm{d}\chi\, W_\mathrm{s}(\chi)\, \delta_\mathrm{s}(\hat{x}, \chi), 
\label{eqn:source_overdensity_field}
\end{equation}
where $\delta_\mathrm{s}$ is the three-dimensional source overdensity and the weighting function is given by the normalized redshift distribution of the tracer,
\begin{equation}
W_\mathrm{s}(\chi) = \frac{n_\mathrm{s}(z(\chi))}{\bar{n}_\mathrm{s}} \frac{\mathrm{d}z}{\mathrm{d}\chi},
\label{eqn:source_wt_fxn}
\end{equation}
with $n_\mathrm{s}(z)$ the redshift distribution of sources and $\bar{n}_s$ their mean angular number density. This kernel applies to galaxies, clusters, superclusters, and radio sources, each using their respective redshift distributions. 

For the power spectrum calculation, we treat each tracer differently according to the complexity of its galaxy-halo connection. For DESI LRGs, the halo occupation distribution (HOD) is well studied (see Equation~4 and 5 of \citet{2024MNRAS.530..947Y} for reference). We model the mean number of central and satellite LRGs per halo using the standard five-parameter HOD framework~\citep{2002ApJ...575..587B, 2005ApJ...633..791Z, 2007ApJ...667..760Z, 2023OJAp....6E..39A} and fix parameters to the median of posteriors from clustering analysis of the same DESI LRG sample~\citep{2024MNRAS.530..947Y}.

For X-ray-selected galaxy clusters and superclusters, and LoTSS radio galaxies, we treat each population as a biased tracer of the matter field at large scales, such that the cross-power spectrum between the tracer number density and the electron density field is
\begin{equation}
P_\mathrm{se}(k,z) = b_\mathrm{s}\, P_\mathrm{me}(k,z),
\label{eqn:galaxy_bias}
\end{equation}
where $P_\mathrm{me}(k,z)$ is the matter-electron cross-power spectrum and $b_\mathrm{s}$ is the large-scale halo bias of the tracer. For X-ray clusters and superclusters, we use the \citet{2010ApJ...724..878T} halo bias evaluated at the median halo mass and redshift of the respective samples, and for LoTSS radio galaxies, we adopt a constant linear bias $b_{s_R} = 2.68$, following the measurement from a similar flux-limited radio source sample~\citep{2024MNRAS.527.6540H, 2024A&A...681A.105N}. To compute the cross-power spectrum with matter field, the total matter profile is decomposed into three components: central galaxy ($\rho_\mathrm{cga}$), gas ($\rho_\mathrm{gas}$), and collisionless matter ($\rho_\mathrm{clm}$), the last of which includes contributions from both, dark matter and satellite galaxies:
\begin{equation}
\rho_\mathrm{m}(r|M) = \rho_\mathrm{cga}(r|M) + \rho_\mathrm{gas}(r|M) + \rho_\mathrm{clm}(r|M).
\label{eqn:matter_profile}
\end{equation}
The central galaxy profile is modeled as a point mass at the halo center, where its halo mass fraction is given by
\begin{equation}
f_\mathrm{cga}(M) = 2A \left[\left(\frac{M}{M_1}\right)^{\tau + \tau_\delta} + \left(\frac{M}{M_1}\right)^{\eta + \eta_\delta}\right]^{-1},
\label{eqn:f_cga}
\end{equation}
with power law indices \{$\tau_\delta$, $\eta_\delta$\} governing the partition of stellar mass between the central and satellite galaxies. The collisionless matter profile starts from a truncated NFW profile and is adiabatically relaxed in response to the redistribution of baryons, by mapping an initial sphere of radius $r_i$ with mass $M_i = M_\mathrm{tNFW}(<r_i)$ to a final radius $r_f$ with mass $M_f = f_\mathrm{clm}(M_i) M_\mathrm{tNFW}(<r_i) + M_\mathrm{gas}(<r_f) + M_\mathrm{cga}(<r_f)$ via
\begin{equation}
\frac{r_f}{r_i} = 1 + a_\psi\left[\left(\frac{M_i}{M_f}\right)^{n_\psi} - 1\right],
\label{eqn:adiabatic_clm_expansion}
\end{equation}
where the mass fraction of collisionless matter is
\begin{equation}
f_\mathrm{clm}(M) = 1 - \frac{\Omega_{\mathrm{b}}}{\Omega_{\mathrm{m}}} + f_\mathrm{sga}(M),
\label{eqn:f_clm}
\end{equation}
and the satellite galaxy mass fraction is $f_\mathrm{sga}(M) = f_\mathrm{star}(M) - f_\mathrm{cga}(M)$. We fix $\tau_\delta=0$, $a_\psi=0.3$ and $n_\psi=2$, following standard practices in baryonification literature~\citep{2015JCAP...12..049S, 2019JCAP...03..020S, 2021JCAP...12..046G}.

\subsubsection{Galaxy and CMB Weak Gravitational Lensing}

The lensing convergence field $\kappa(\hat{x})$ quantifies the projected mass overdensity along the line-of-sight,
\begin{equation}
\kappa(\hat{x}) = \int\limits_0^{\chi_\mathrm{H}} \mathrm{d}\chi W_\gamma(\chi) \delta_\mathrm{m}(\hat{x}, \chi),
\label{eqn:convergence_field}
\end{equation}
where $\delta_\mathrm{m}$ is the matter overdensity and the lensing efficiency kernel is
\begin{equation}
W_\gamma(\chi) = \frac{3H_0^2\Omega_\mathrm{m}}{2c^2} \chi (1+z(\chi)) \int\limits_\chi^{\infty} \mathrm{d}\chi^\prime \frac{n_\mathrm{s}(\chi^\prime)}{\bar{n}_\mathrm{s}} \frac{\chi^\prime - \chi}{\chi^\prime},
\label{eqn:lensing_wt_fxn}
\end{equation}
where $n_\mathrm{s}(\chi^\prime) = n_\mathrm{s}(z)\, \mathrm{d}z/\mathrm{d}\chi^\prime$ and $\bar{n}_\mathrm{s}$ denote the redshift distribution and mean angular number density of source galaxies, respectively. For CMB lensing, the source plane is fixed at the surface of last scattering ($z_\mathrm{CMB} \sim 1100$), so the source redshift distribution is replaced by a Dirac delta function at $z_\mathrm{CMB}$, yielding the CMB lensing kernel,
\begin{equation}
W_{\kappa_\mathrm{CMB}}(\chi) = \frac{3H_0^2\Omega_\mathrm{m}}{2c^2} \chi (1+z(\chi)) \frac{\chi(z_\mathrm{CMB}) - \chi}{\chi(z_\mathrm{CMB})}.
\label{eqn:CMB_lensing_wt_fxn}
\end{equation}
The cross-power spectrum between the matter and electron overdensity fields, $P_\mathrm{me}(k, z)$, is computed using the matter and gas density profiles, as defined in Equations~\ref{eqn:matter_profile} and \ref{eqn:gas_profile}, respectively.

\subsubsection{Cosmic Infrared Background}

The observed CIB intensity along a line-of-sight $\hat{x}$ is
\begin{equation}
    I_\nu(\hat{x}) = \int\limits_0^{\chi_\mathrm{H}} \mathrm{d}\chi\, W_\mathrm{CIB}(\chi)\, \delta_\mathrm{CIB}(\hat{x}, \chi),
\end{equation}
where the CIB weighting kernel is
\begin{equation}
    W_\mathrm{CIB}(\chi) = \dfrac{\mathrm{d}I_\nu}{\mathrm{d}z} \frac{\mathrm{d}z}{\mathrm{d}\chi},
\end{equation}
with ${\mathrm{d}I_\nu}/{\mathrm{d}z}$ the tomographic extragalactic background light intensity at frequency $\nu$. The overdensity of sources contributing to CIB is defined as $\delta_\mathrm{CIB} = b_\mathrm{CIB} \delta_\mathrm{m}$, where $b_\mathrm{CIB}(z)$ is the effective linear clustering bias. We use the ${\mathrm{d}I_\nu}/{\mathrm{d}z}$ and $b_\mathrm{CIB}(z) {\mathrm{d}I_\nu}/{\mathrm{d}z}$ models fitted to direct redshift tomographic cross-correlation measurements from \citet{2025ApJ...992...65C}.

\subsubsection{Thermal Sunyaev-Zel'dovich Effect}

The Compton-$y$ parameter quantifies the line-of-sight integral of the electron thermal pressure $\mathcal{P}_\mathrm{e}$, 
\begin{equation}
y(\hat{x}) = \int\limits_0^{\chi_\mathrm{H}} \mathrm{d}\chi\, W_y(\chi)\, \mathcal{P}_\mathrm{e}(\hat{x}, \chi),
\label{eqn:compton_y}
\end{equation}
where the tSZ effect weighting kernel is
\begin{equation}
W_y(\chi) = \frac{\sigma_\mathrm{T}}{m_\mathrm{e} c^2} \frac{1}{(1+z)},
\label{eqn:tSZ_wt_fxn}
\end{equation}
with $\sigma_\mathrm{T}$ the Thomson scattering cross-section, and $m_\mathrm{e}$ the electron mass. The angular cross-correlation between the Compton-$y$ parameter and the electron overdensity field depends on the cross-power spectrum of electron thermal pressure and gas density, $P_\mathrm{pe}(k,z)$. Assuming hydrostatic equilibrium, the total pressure profile of a halo is calculated by solving,
\begin{equation}
\frac{1}{\rho_\mathrm{gas}(r)}\frac{\mathrm{d}\mathcal{P}_\mathrm{tot}}{\mathrm{d}r} = -\frac{GM(<r)}{r^2}.
\label{eqn:hydrostatic_equilibrium}
\end{equation}
The thermal pressure profile is then written as $\mathcal{P}_\mathrm{th} = \mathcal{P}_\mathrm{tot} (1-R_\mathrm{nt})$, where the non-thermal pressure fraction $R_\mathrm{nt}$ is parameterized as
\begin{equation}
R_\mathrm{nt} = \frac{\mathcal{P}_\mathrm{nt}} {\mathcal{P}_\mathrm{tot}} = \alpha_\mathrm{nt}\, f(z) \left(\frac{r}{r_{200c}}\right)^{\gamma_\mathrm{nt}},
\label{eqn:non_thermal_pressure_fraction}
\end{equation}
with $\alpha_\mathrm{nt}=0.3$ and $\gamma_\mathrm{nt}=0.3$ controlling the amplitude and the slope~\citep{2023MNRAS.519.2069O, 2010ApJ...725.1452S}. The redshift evolution function~\citep[$\nu_\mathrm{nt} = 0.3$;][]{2010ApJ...725.1452S} is
\begin{equation}
f(z) = \min\left[(1+z)^{\nu_\mathrm{nt}},\,\left(\frac{6^{-\gamma_\mathrm{nt}}}{\alpha_\mathrm{nt}} - 1\right)\tanh(\nu_\mathrm{nt} z) + 1\right].
\label{eqn:non_thermal_pressure_fraction_redshift_evolution}
\end{equation}
Finally, the electron thermal pressure is computed as
\begin{equation}
\mathcal{P}_\mathrm{e} = \frac{2X_\mathrm{H}+2}{5X_\mathrm{H}+3}\mathcal{P}_\mathrm{th},
\label{eqn:electron_thermal_pressure}
\end{equation}
assuming hydrogen mass fraction $X_\mathrm{H} = 0.75$.

\subsubsection{X-ray Count Rates}

The projected X-ray count rate density is given by,
\begin{equation}
C(\hat{x}) = \int\limits_0^{\chi_\mathrm{H}} \mathrm{d} \chi \, W_X(\chi) \, c_X(\hat{x}, \chi),
\label{eqn:count_rate_density_field}
\end{equation}
where $c_X$ is the three-dimensional X-ray photon count rate density and $W_X$ is the X-ray weighting function~\citep{2024PhRvL.133e1001F,2025PhRvD.112d3525L},
\begin{equation}
W_X(\chi) = \frac{1}{4\pi} \frac{1}{(1+z)^3}.
\label{eqn:Xray_wt_fxn}
\end{equation}
The halo count rate density profile~\citep{2023MNRAS.518.1496L, 2024PhRvL.133e1001F,2025PhRvD.112d3525L} is
\begin{equation}
c_X(r, z) = n_\mathrm{H}(r)\, n_\mathrm{e}(r) \int\limits_{E_\mathrm{min}(1+z)}^{E_\mathrm{max}(1+z)} \Lambda(T(r), Z(r), E) \, A(E) \, \mathrm{d}E,
\label{eqn:count_rate_profile}
\end{equation}
where the comoving number densities of free electrons and protons are calculated as $\rho_\mathrm{gas}(1+X_\mathrm{H})/2m_\mathrm{p}$ and $\rho_\mathrm{gas}X_\mathrm{H}/m_\mathrm{p}$, respectively, and the electron temperature is computed using the ideal gas law as $T = \mathcal{P}_\mathrm{e}/n_\mathrm{e}k_\mathrm{B}$, with $k_\mathrm{B}$ the Boltzmann constant. We use \textsc{APEC} tables~\citep{2001ApJ...556L..91S} to compute the X-ray cooling function $\Lambda$, integrated over the X-ray energy band [$E_\mathrm{min}, E_\mathrm{max}$] in observer's frame, and assume empirical metallicity profiles, $Z(r)$~\citep{2018SSRv..214..129M}. This band-limited cooling function includes line emission, continuum emission from Bremsstrahlung and radiative recombination, and pseudo continuum emission from weak, unresolved spectral lines. We construct the ROSAT instrumental response using the redistribution matrix (RMF/RSP) and the auxiliary response (ARF), combining them to form the energy redistribution matrix weighted by the effective area. The bandpass (effective area as a function of input energy, $A(E)$) is then obtained by summing this matrix over detector channels within the chosen energy range ($E_\mathrm{min}, E_\mathrm{max}$). This procedure maps incident photon energies to detected counts while fully accounting for the detector response.

\subsection{Cross-Correlation Estimators} \label{subsec:cross_correlation_estimators}

We measure the angular cross-correlation between FRB DMs and each of the tracers described in Section~\ref{sec:data} using the \textsc{treecorr} package~\citep{2015ascl.soft08007J}. The FRB DMs are constructed as a scalar quantity, $\mathcal{D}_i = \mathrm{DM}_i - \langle \mathrm{DM} \rangle$, assigned to each FRB position, where $\langle \mathrm{DM} \rangle$ is the mean DM computed over all FRBs passing the DM threshold within the footprint of the corresponding tracer. Each measurement is conducted with multiple lower DM threshold cuts, DM$_\mathrm{cut}$, retaining only FRBs with DM $> \mathrm{DM}_\mathrm{cut}$. We use DM$_\mathrm{cut} \in \{500,~750,~1000\}$~pc\,cm$^{-3}$. The signal is not expected to grow monotonically with the threshold: each cut represents a different trade-off between sample size, noise, and redshift sensitivity, and the response of the cross-correlation to this selection depends on the radial kernel of each tracer. Since FRB redshifts are unknown for our sample, the DM cut serves as a proxy for redshift depth, and reporting results across multiple thresholds serves as a consistency check, verifying that the measured signal is stable across selections and not driven by any particular choice of threshold.

The estimator of angular cross-correlation with a (discrete) disperser catalog $A$, $\hat{\xi}^{A\mathcal{D}}(\theta)$, quantifies the DM excess at angular separation $\theta$ around the disperser relative to random locations, and is written as
\begin{equation}
\begin{aligned}
\hat{\xi}^{A\mathcal{D}}(\theta) = \frac{\sum\limits_{i \in \mathrm{dis}} \sum\limits_{j \in \mathrm{FRB}} w_{ij} \mathcal{D}_j \, \Theta_{ij}}{\sum\limits_{i \in \mathrm{dis}} \sum\limits_{j \in \mathrm{FRB}} w_{ij} \Theta_{ij}} - \frac{\sum\limits_{i \in \mathrm{rand}} \sum\limits_{j \in \mathrm{FRB}} w_{ij} \mathcal{D}_j \, \Theta_{ij}}{\sum\limits_{i \in \mathrm{rand}} \sum\limits_{j \in \mathrm{FRB}} w_{ij} \Theta_{ij}},
\label{eqn:NK_estimator}
\end{aligned}
\end{equation}
where $\Theta_{ij} = \Theta(\theta_{ij}-\theta)$ selects pairs of disperser indexed $i$ and FRB indexed $j$ with angular separation $\theta_{ij}$ falling in angular bin centered at $\theta$, and $w_{ij}$ are the product of per-disperser and per-FRB weights. Unless otherwise stated, the weights $w_{ij}$ are fixed to unity. The randoms$\times$FRB pairs gives the null expectation, where the disperser positions are randomly distributed; removing this corrects for survey geometry or footprint effects, leaving the signal due to true disperser-DM correlations. The physical space estimator is constructed using the \textsc{rlens} metric when disperser redshifts are available.

The estimator $\hat{\xi}^{\mathcal{BD}}(\theta)$ for angular cross-correlation with a continuous scalar field $B$ quantifies the DM fluctuation at angular separation $\theta$ from regions of tracer field fluctuations $\mathcal{B} = B - \langle B \rangle$,
\begin{equation}
\hat{\xi}^{\mathcal{BD}}(\theta) = \dfrac{\sum\limits_{p \in B} \sum\limits_{j \in \mathrm{FRB}} w_{ij} \mathcal{B}_p \, \mathcal{D}_j \, \Theta_{jp}}{\sum\limits_{p \in B} \sum\limits_{j \in \mathrm{FRB}} w_{ij} \Theta_{jp}},
\label{eqn:KK_estimator}
\end{equation}
where $\langle B \rangle$ is the field average over the overlapping survey footprint and $\mathcal{B}_p$ denotes the field evaluated at pixel $p$. The mean subtraction for continuous tracer fields implies that we are measuring the correlation between fluctuations in both fields.

The cross-correlation estimator for galaxy weak lensing, $\hat{\xi}^{\gamma\mathcal{D}}(\theta)$, is written as
\begin{equation}
\hat{\xi}^{\gamma\mathcal{D}}(\theta) = \frac{\sum\limits_{i \in \mathrm{source}} \sum\limits_{j \in \mathrm{FRB}} w_{ij} \mathcal{D}_j \gamma_{ij} \Theta_{ij}}{\sum\limits_{i \in \mathrm{source}} \sum\limits_{j \in \mathrm{FRB}} w_{ij} \Theta_{ij}},
\label{eqn:GK_estimator}
\end{equation}
where $\gamma_{ij}$ is the tangential shear of source galaxy $i$ measured with respect to the direction towards FRB $j$. A positive signal $\hat{\xi}^{\gamma\mathcal{D}}>0$ indicates that FRBs with higher DMs are preferentially located behind regions of higher projected mass density, as the convergence field that produces the tangential shear is also correlated with the gas that contributes to the FRB DM. Since the shear catalog is provided as per-galaxy ellipticity components ($g_{1,i}, g_{2,i}$), \textsc{treecorr} internally computes the tangential shear component, $\gamma$, for each FRB-galaxy pair by rotating the complex ellipticity into the frame defined by the separation vector between the pair.

We assess systematic uncertainty from small sample fluctuations using a null test constructed by randomly shuffling FRB DMs while keeping their sky positions fixed, thereby removing correlations with LSS (see Appendix~\ref{sec:null_tests}). Across 100 such realizations, the measured signal is consistent with the null expectation, indicating no evidence for spurious correlations (see Figure~\ref{fig:null_tests_summary_onlyDMshuffled}).

\subsection{Jackknife Covariance Estimation} \label{subsec:jackknife_covariance_estimation}

Covariance for all cross-correlation measurements are estimated using a spatial jackknife procedure. The survey footprint of each probe is divided into $N_\mathrm{patch}$ spatially contiguous patches using the $k$-means patch assignment algorithm, which partitions sky into patches of approximately equal area. For each jackknife realization, one patch is omitted from both the tracer and the FRB catalogue simultaneously, and the cross-correlation is recomputed on the remaining sky. The jackknife covariance is then
\begin{equation}
\hat{C}_{ab} = \dfrac{N_\mathrm{patch}-1}{N_\mathrm{patch}} \sum\limits_{k=1}^{N_\mathrm{patch}} \left( \hat{\xi}_a^k - \bar{\xi}_a \right) \left( \hat{\xi}_b^k - \bar{\xi}_b \right),
\label{eqn:jackknife_covariance}
\end{equation}
where $\hat{\xi}_a^k$ is the correlation function in separation bin $a$ with patch $k$ omitted, and $\bar{\xi}_a$ is the mean over all jackknife realizations. We use $N_\mathrm{patch} = 50$ for all probes, except for cross-correlation with DECADE weak lensing, where we use $N_\mathrm{patch} = 40$ because of the poor overlap between DECADE and CHIME footprints. This is done to ensure that the number of FRBs within the footprint exceeds the number of patches at all DM thresholds.

To jointly characterize the covariance structure across multiple DM threshold cuts, which are essentially nested subsets of the full FRB sample, we construct a joint data vector, $\hat{\boldsymbol{\xi}}$, by concatenating the cross-correlation measurements across all separation bins and all DM thresholds for a given probe and estimate the corresponding joint covariance matrix using multi-estimator jackknife covariance procedure. This joint covariance, $\hat{\mathbf{C}}$, captures both, the bin-to-bin correlations within a single DM cut and the correlations between the measurements at different DM cuts arising from the overlap between FRB samples selected at different thresholds. To reduce stochastic noise in the jackknife covariance arising from the arbitrary choice of patch boundaries, we further average the covariance estimates using 100 independent realizations of the patch assignment. Using this measured covariance, we also report the correlation matrix
\begin{equation}
r_{ij} = \dfrac{\hat{\mathbf{C}}_{ij}}{\sqrt{\hat{\mathbf{C}}_{ii} \hat{\mathbf{C}}_{jj}}},
\label{eqn:correlation_coefficient}
\end{equation}
where $r_{ij}$ is the correlation coefficient and \{$i$, $j$\} are the corresponding indices of the covariance matrix. 

Since the covariance matrix $\hat{\mathbf{C}}$ is estimated from a finite number of jackknife patches, we expect this covariance to be inherently noisy. While the sample covariance is an unbiased estimator of the true covariance, its inverse is biased, which directly propagates into any quantity that depends on its inverse, such as the signal-to-noise ratio. We therefore use the \citet{2007A&A...464..399H} factor to define the unbiased inverse covariance,
\begin{equation}
\hat{\mathbf{C}}^{-1}_\ast = \frac{N_\mathrm{patch} - N_\mathrm{data} - 2}{N_\mathrm{patch} - 1} \hat{\mathbf{C}}^{-1},
\label{eqn:Hartlap_factor}
\end{equation}
where $N_\mathrm{data}$ is the length of the data vector. The Hartlap multiplicative correction removes the bias under the assumption of Gaussian-distributed data vectors, thus ensuring that the statistical inferences are neither artificially overconfident, nor biased.

\begin{table*}
    \centering
    \begin{tabular}{lrccc}
        \toprule
        Tracer & Statistic & $N_\mathrm{FRB}$ & SNR & SNR$_\mathrm{null}$ \\
        \midrule

        DESI LRG & $\xi^{\mathrm{g}\mathcal{D}}(R)$ & 1121/611/323 & 2.8 & 0.26 \\

        DECADE Weak Lensing & $\xi^{\gamma\mathcal{D}}(\theta)$ & 196/97/43 & 2.6 & 0.26 \\

        100~$\mu$m Cosmic Infrared Background & $\xi^{I_\mathrm{CIB}\mathcal{D}}(\theta)$ & 942/518/286 & 4.0 & 0.31 \\

        Planck CMB lensing convergence & $\xi^{\kappa_\mathrm{CMB}\mathcal{D}}(\theta)$ & 1300/700/378 & 3.3 & 0.31 \\

        Planck Compton-$y$ & $\xi^{y\mathcal{D}}(\theta)$ & 1148/629/339 & 3.8 & 0.28 \\

        X-ray groups and clusters (RASS) & $\xi^{c\mathcal{D}}(R)$ &  706/393/217 & 5.0 & 0.33 \\

        ~~~~~~~~~~~~~~~~~~~~~~~~~~~~~~~~~~ (eRASS1) & &  150/81/41 & 3.2 & 0.31 \\

        eRASS1 X-ray superclusters &  $\xi^{sc\mathcal{D}}(R)$ & 55/32 & 3.3 & 0.14 \\

        RASS soft Xray background & $\xi^{C_X\mathcal{D}}(\theta)$ & 379/223/129 & 4.1 & 0.36 \\

        LoTSS radio sky flux density & $\xi^{S_\nu\mathcal{D}}(\theta)$ & 1336/729/391 & 3.0 & 0.35 \\

        LoTSS radio sources & $\xi^{s_R\mathcal{D}}(\theta)$ & 1741/964/528 & 2.4 & 0.22 \\
        
        \midrule
    \end{tabular}
    
    \caption{Summary of cross-correlation measurement statistics for each of the ten tracers. The statistic column lists the cross-correlation estimator used for each tracer, where $\mathcal{D}$ denotes the FRB DM field and superscripts identify the tracer: $g$ (galaxy number density), $\gamma$ (tangential shear), $I_\mathrm{CIB}$ (CIB intensity), $\kappa_\mathrm{CMB}$ (CMB lensing convergence), $y$ (Compton-$y$ parameter), $c$ (X-ray groups and clusters), $sc$ (X-ray superclusters), $C_X$ (soft X-ray background), $S_\nu$ (radio flux density), and $s_R$ (radio source number density). $N_\mathrm{FRB}$ gives the number of FRBs contributing to each measurement for respective DM cuts. The measurement and null SNR is computed from the joint data vectors across all three DM cuts and their jackknife cross-covariance matrix, fully accounting for the covariance between nested DM subsamples.}
    \label{table:detection_statistics}
\end{table*}

\subsection{Signal-to-Noise Estimation} \label{subsec:SNR_estimation}

The signal-to-noise ratio (SNR) for cross-correlation with each probe is computed from the joint data vector $\hat{\boldsymbol{\xi}}$, and the corresponding unbiased inverse covariance matrix $\hat{\mathbf{C}}^{-1}_\ast$, as defined in Section~\ref{subsec:jackknife_covariance_estimation}. The $\chi^2$ for null hypothesis (no-correlation) is defined as
\begin{equation}
\chi^2 = \hat{\boldsymbol{\xi}}^T \hat{\mathbf{C}}^{-1}_\ast \hat{\boldsymbol{\xi}},
\label{eqn:chi_sq}
\end{equation}
using which the measurement SNR is computed following \citet{2022PhRvD.105j3537S}. This SNR accounts for the total variance, including cosmic variance and shot noise.

\section{Measured Correlation Functions} \label{sec:measured_correlation_functions}

\begin{figure*}[ht!]
\centering
\includegraphics[width=\textwidth]{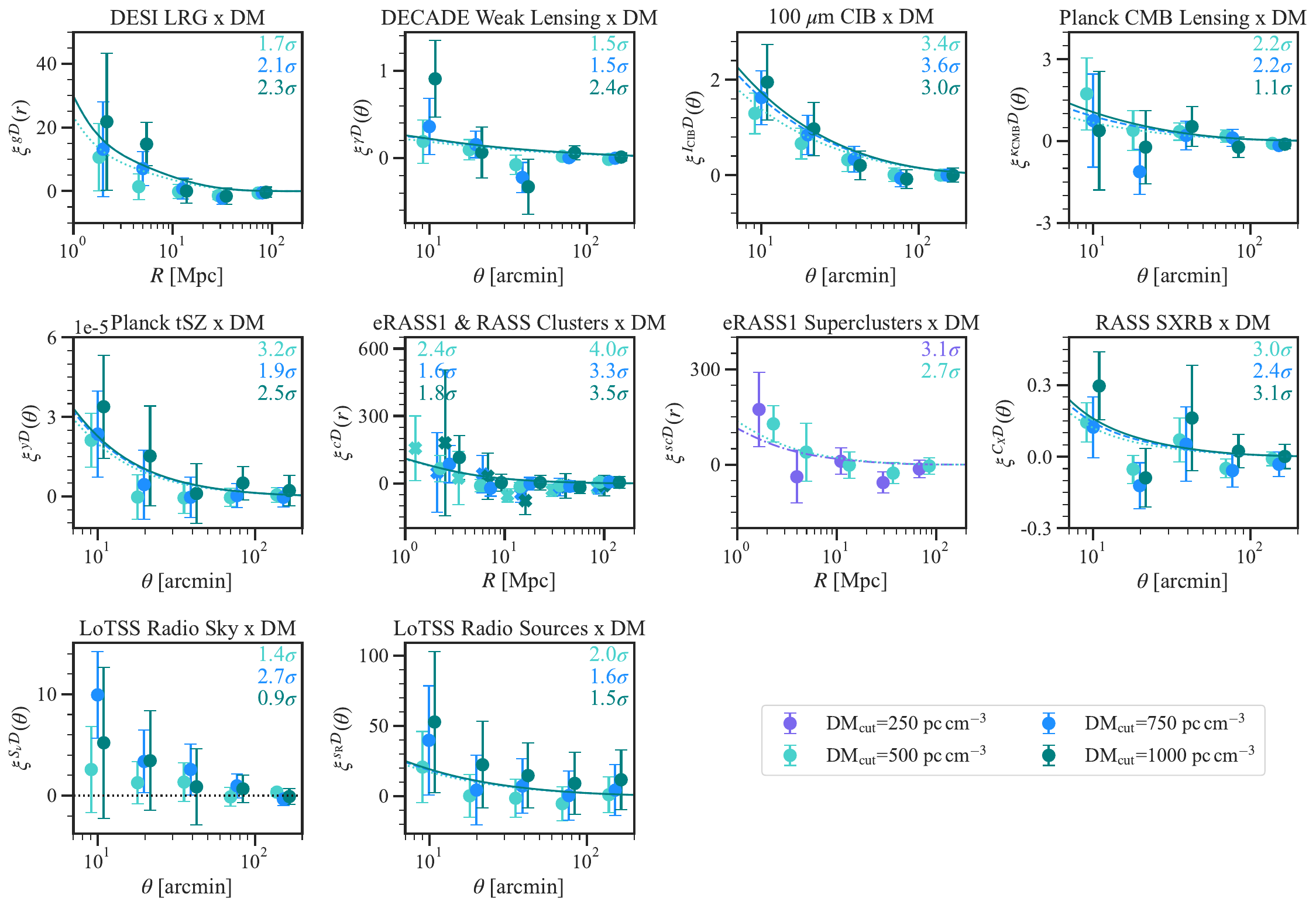}
\caption{Correlation functions between CHIME/FRB DMs and each of the ten tracers, measured using \textsc{treecorr} with jackknife covariance estimation over 50 spatial patches. Correlations are measured as a function of projected physical separation $R$ at $\gtrsim$ Mpc-scales, or angular separation $\theta$ at $\gtrsim$ arcminute/degree-scales, depending on the tracer. Each panel shows results for three DM thresholds, retaining only FRBs with DM$_\mathrm{ex} \geq$ DM$_\mathrm{cut}$. The measurement significance quoted in each panel is the SNR for each DM cut; joint SNRs across all DM cuts are listed in Table~\ref{table:detection_statistics}. Where available, theoretical predictions computed using a \textsc{BCEmu} baryonic model are overlaid for reference.}
\label{fig:FRB_cross_correlations}
\end{figure*}

\begin{figure*}[ht!]
\centering
\includegraphics[width=\textwidth]{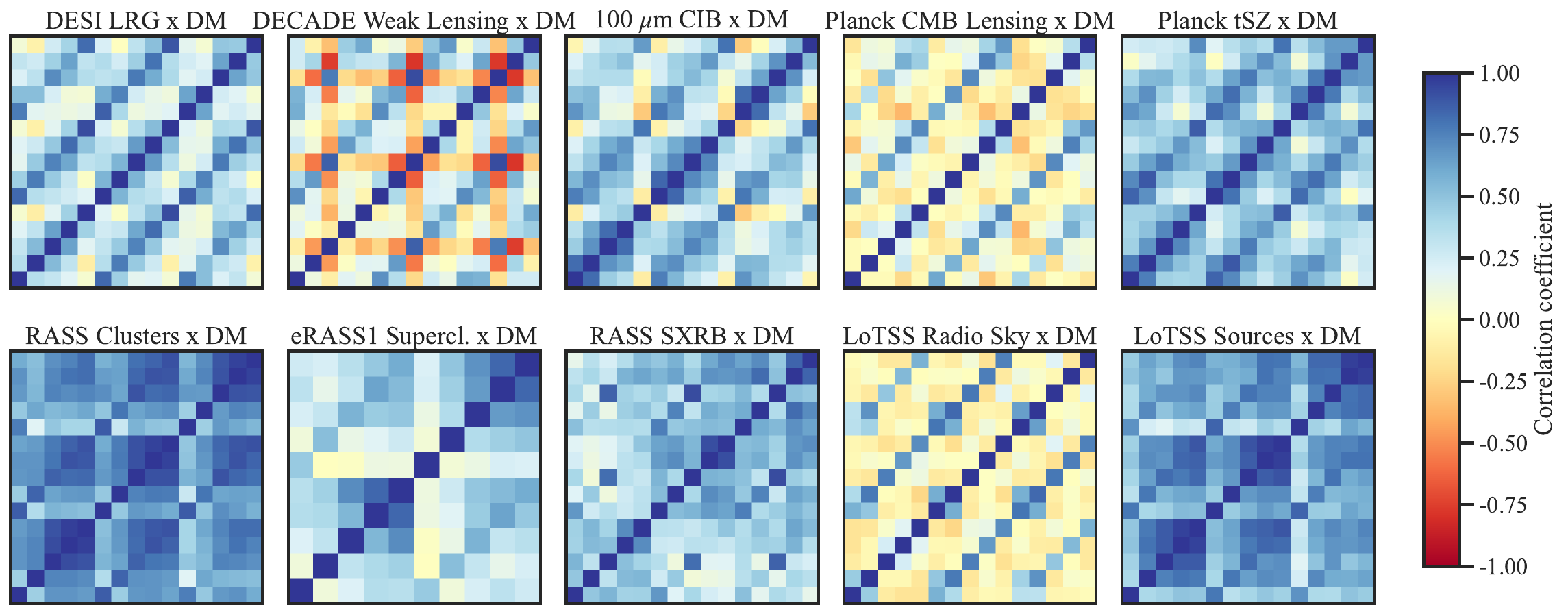}
\caption{Correlation matrices from the jackknife covariance between angular bins for each of the ten correlation functions, computed jointly across the three DM threshold samples. Each matrix element shows the Pearson correlation coefficient $C_{ij}/\sqrt{C_{ii}C_{jj}}$. The block structure within each matrix reflects correlations between separation bins at fixed DM cut, while the off-diagonal blocks capture correlations between measurements at different DM thresholds, which arise naturally because higher DM cut samples are nested subsets of lower DM cut samples. The degree of off-diagonal structure varies across tracers, reflecting differences in the underlying angular clustering of each tracer field. Matrices dominated by a clean diagonal (e.g. Planck CMB lensing, LoTSS radio sky) indicate largely uncorrelated bins, while those with significant off-diagonal power (e.g. DESI LRG, DECADE weak lensing) indicate stronger inter-bin correlations that must be accounted for in the joint SNR estimate.}
\label{fig:correlation_matrices}
\end{figure*}

We present the measured angular cross-correlation functions between the CHIME/FRB Catalog 2 DM field and each of the ten LSS and baryon tracers described in Section~\ref{sec:data}. The physical-space and angular-space correlations are measured in five logarithmically spaced bins spanning $1 \leq R \leq 100$~Mpc and $10^\prime \leq \theta \leq 200^\prime$, respectively. The lower limit of separation bins is restricted by the FRB localization uncertainties and the beam size of surveys used. The full set of correlation functions is shown in Figure~\ref{fig:FRB_cross_correlations}, with each panel corresponding to a single tracer and displaying results across all DM thresholds. Theoretical predictions from the \textsc{BCEmu} baryonic model are overlaid where available. A summary of the measurement statistics, including the estimator, the number of contributing FRBs at each DM threshold, and SNRs of the measurement and the null expectation (SNR$_\mathrm{null}$, see Appendix~\ref{sec:null_tests}), is provided in Table~\ref{table:detection_statistics}. The correlation matrices, computed using the jackknife covariances estimated jointly across all DM thresholds, are displayed in Figure~\ref{fig:correlation_matrices}. The off-diagonal structure in these matrices reflects the nested overlap between FRB sub-samples at different DM thresholds, which are fully accounted for in the joint SNR estimate. We organize the results by tracer type below.

\subsection{Large-Scale Structure Tracers} \label{subsec:LSS_tracers}

We present the cross-correlation measurements between the CHIME/FRB DM field and the four LSS tracers described in Sections~\ref{subsubsec:DESI_LRGs}, \ref{subsubsec:DECADE_WL}, \ref{subsubsec:WISE_CIB}, and \ref{subsubsec:Planck_CMB_lensing} with complementary redshift sensitivities (see Figure~\ref{fig:redshift_sensitivity}): DESI LRGs - $z \in (0.4, 0.5)$, DECADE weak lensing - $z \in (0.2, 0.7)$, WISE-reconstructed 100~$\mu$m CIB intensity - $z \in (0.2, 0.8)$, and Planck CMB lensing convergence - $z \in (0.4, 1.1)$.

We begin with the projected physical-space angular correlation function $\hat{\xi}^{g\mathcal{D}}(R)$ between the DESI LRG sample in the first tomography bin and the FRB DMs. The results are shown in the DESI LRG panel of Figure~\ref{fig:FRB_cross_correlations}, with a positive signal indicating that FRB sightlines passing through regions of higher projected LRG overdensity carry systematically larger DMs. The per-DM-cut measurement significances are $1.7\sigma$, $2.1\sigma$, and $2.3\sigma$ for DM$_\mathrm{cut} \in \{500,~750,~1000\}$~pc\,cm$^{-3}$, combining to a joint SNR of $2.8\sigma$. Theory curves computed by fixing HOD parameters to the posteriors of \citet{2024MNRAS.530..947Y} are in broad agreement with the measured signal.

Moving to weak lensing, we cross-correlate the FRB DM field with DECADE source galaxy shapes in the fourth tomography bin using the tangential shear estimator $\hat{\xi}^{\gamma \mathcal{D}}(\theta)$. For each galaxy, we use the inverse-variance weights in our calculation, as provided in the DECADE catalog. A positive signal, indicating that FRBs with higher extragalactic DMs are preferentially located behind regions of higher projected mass density, is measured with per-DM-cut significances of $1.6\sigma$, $1.5\sigma$, and $2.4\sigma$ for the same DM thresholds, combining to a joint SNR of $2.6\sigma$. The theoretical predictions are in broad agreement with the measurement.

Next, we cross-correlate the DM field with WISE-reconstructed 100~$\mu$m CIB intensity map, measuring the angular correlation function $\hat{\xi}^{I_\mathrm{CIB} \mathcal{D}}(\theta)$. A positive signal, indicating that FRBs with higher extragalactic DMs are preferentially located behind regions of higher projected CIB intensity, is measured with per-DM-cut significances of $3.4\sigma$, $3.6\sigma$, and $3.0\sigma$ for aforementioned DM thresholds, combining to a joint SNR of $4.0\sigma$. While the signal is expected to be dominated by the IGM, the correlation of host galaxy DM contribution with CIB may contribute to the measured signal since FRB sources have strong ties to star-formation~\citep{2023ApJ...954...80G, 2024Natur.635...61S}, and is worth exploring in future.

Extending the lensing cross-correlation to CMB source plane, we measure the correlation function, $\hat{\xi}^{\kappa_\mathrm{CMB}\mathcal{D}}(\theta)$, with per-DM-cut measurement significances of $2.2\sigma$, $2.2\sigma$, and $1.1\sigma$ for aforementioned DM thresholds, combining to a joint SNR of $3.3\sigma$. Each FRB was weighted by the value of lensing kernel at our approximate FRB redshift from mean DM-$z$ relation. We note that while higher DM thresholds preferentially select more distant FRBs with greater cosmological depth, and therefore increase overlap with CMB lensing kernel, the per-cut measurement significance does not necessarily increase with DM$_\mathrm{cut}$. The declining significance at larger DM thresholds reflects the reduced FRB sample size (from 1300 for DM$_\mathrm{cut} = 500$~pc\,cm$^{-3}$ to 378 for DM$_\mathrm{cut} = 1000$~pc\,cm$^{-3}$) rather than a loss of cosmological depth; the resulting increase in shot noise can dominate over the gain in signal depth, particularly for tracers whose redshift kernels have substantial overlap with even the lowest DM threshold sample. The measured amplitude and scale dependence of the signal is in broad agreement with our theory model. 

As a sanity check, cross-correlations using the standard TT-only, PP-only and MV reconstructions yield joint SNRs of $3.2\sigma$, $1.3\sigma$, and $3.5\sigma$ respectively, confirming that the TT-only quadratic estimator dominates the signal-to-noise at Planck noise levels, and that the tSZ deprojection does not significantly degrade the measurement. This further confirms that contamination from the tSZ effect, when using a TT-only quadratic estimator without explicit tSZ deprojection, has minimal impact on our measurement. The stability of SNRs between TT, tSZ-deprojected TT, and MV reconstructions also reassures that our measurement is insensitive to small biases from CIB in lensing reconstruction.

\subsection{Thermal and Ionized Gas Tracers} \label{subsec:baryon_tracers}

We present the cross-correlation measurements between the CHIME/FRB DM field and the four thermal and ionized gas tracers described in Section~\ref{subsubsec:Planck_tSZ}-\ref{subsubsec:RASS_clusters_SXRB} with complementary redshift sensitivities, as illustrated in Figure~\ref{fig:redshift_sensitivity}: the Compton $y$-parameter and SXRB are broadly sensitive to baryon distribution at $z \in (0.1, 0.75)$ and  $z \in (0.1, 0.65)$, respectively, while the X-ray cluster and supercluster samples are concentrated at $z \in (0.1, 0.4)$, with redshift distributions peaking at $z \sim 0.2$. Together, they provide a multi-scale view of the hot and warm baryon distribution, spanning from high-pressure ICM of individual galaxy clusters to diffuse gas in filaments and outskirts of LSS.

We begin with the tSZ cross-correlation, measuring the angular cross-correlation function $\hat{\xi}^{y\mathcal{D}}(\theta)$. As shown in Figure~\ref{fig:FRB_cross_correlations}, a positive signal is measured with per-DM-cut measurement significances of $3.2\sigma$, $1.9\sigma$, and $2.5\sigma$ for aforementioned DM thresholds, combining to a joint SNR of $3.8\sigma$. The signal amplitude decreases monotonically with decreasing DM threshold, and is in broad agreement with theoretical expectations. These measurements constrain the pressure-weighted electron distribution along FRB sightlines (see Section~\ref{subsec:feedback_implications}).

To probe the hot gas directly through X-ray emission, and at lower redshifts than the tSZ effect, we cross-correlate the FRB DMs with three complementary X-ray tracers: the RASS~\citep{2023MNRAS.526.3757K} and eRASS1~\citep{2024A&A...685A.106B} cluster catalogs, the eRASS1 supercluster catalog~\citep{2024A&A...683A.130L}, and the RASS SXRB map~\citep{1997ApJ...485..125S, 2024PhRvL.133e1001F}. For the cluster and supercluster catalogs, we measure the physical-space correlation functions $\hat{\xi}^{c\mathcal{D}}(R)$ and $\hat{\xi}^{sc\mathcal{D}}(R)$, with clusters weighted by their halo mass and superclusters weighted by their multiplicity. For the diffuse SXRB, we instead measure the angular correlation function $\hat{\xi}^{C_X\mathcal{D}}(\theta)$. All X-ray cross-correlations are measured at the three standard DM thresholds, DM$_\mathrm{cut} \in \{500,~750,~1000\}$~pc\,cm$^{-3}$, except for superclusters where the limited sample size precludes the use of the highest DM threshold bins.

The results for all X-ray tracers are shown in Figure~\ref{fig:FRB_cross_correlations}, with RASS (circles) and eRASS1 (crosses) results displayed together in a single panel, slightly offset in $r$ for clarity. Consistent with the tSZ measurement, all X-ray tracers yield positive correlations, confirming that FRB sightlines passing through overdense, X-ray luminous environments accumulate systematically larger DMs. The RASS cluster correlation yields the most significant measurement among the X-ray tracers, with per-DM-cut measurement significances of $4.0\sigma$, $3.3\sigma$, and $3.5\sigma$ for aforementioned DM thresholds, combining to a joint SNR of $5.0\sigma$. The eRASS1 cluster correlation yields a consistent, but somewhat lower significance, with per-DM-cut measurements of $2.4\sigma$, $1.6\sigma$, and $1.8\sigma$ for the same DM thresholds, combining to a joint SNR of $3.2\sigma$; the reduced significance relative to RASS primarily reflects the smaller sky overlap with eRASS1 footprint. Extending to the largest physical scales probed in this section, the eRASS1 supercluster cross-correlation yields per-DM-cut measurements of $3.1\sigma$ and $2.7\sigma$ for DM$_\mathrm{cut} = 250$ and $500$~pc\,cm$^{-3}$, respectively, combining to a joint SNR of $3.3\sigma$. Finally, the RASS SXRB cross-correlation, $\hat{\xi}^{C_X\mathcal{D}}(\theta)$, which traces the integrated emission from diffuse gas rather than resolved cluster environments, yields per-DM-cut measurements of $3.0\sigma$, $2.4\sigma$, and $3.1\sigma$, combining to a joint SNR of $4.1\sigma$. Taken together, these measurements span both, the discrete high-density ICM environments of individual galaxy clusters and the diffuse integrated emission from LSS, providing a coherent multi-tracer picture of the hot baryon contribution to the FRB DM budget that is fully consistent with the lensing and tSZ measurements presented above.

\subsection{Radio Continuum Tracers} \label{subsec:Radio_continuum_tracers}

We present the cross-correlation measurements between the CHIME/FRB DM field and the two LoTSS DR3 radio continuum tracers described in Section~\ref{subsubsec:LoTSS_radio_sky_sources}: the 144~MHz continuum flux density map and the flux density-limited discrete source catalog. Both tracers, when combined with FRBs, probe baryon distribution around a combination of star-forming galaxies and AGNs at redshifts $z \sim 0.1-0.8$~\citep{2024MNRAS.527.6540H}. While the continuum flux density map traces the integrated synchrotron emission of the radio sky, receiving contributions from both, resolved point sources and any diffuse emission associated with LSS, the discrete source catalog directly traces individual radio galaxies, making the two measurements complementary probes of the same underlying matter distribution. The minimum angular scale used for the measurement safely avoids the $\lesssim 0.03^\circ$ regime, where multi-component radio sources produce spurious excess clustering~\citep{2024MNRAS.527.6540H}, and the chosen maximum angular scale limits us to the regime where angular clustering of LoTSS sources is well described by linear bias models.

The results for both tracers, shown in Figure~\ref{fig:FRB_cross_correlations}, are consistent with the LSS measurements presented in Section~\ref{subsec:LSS_tracers} and \ref{subsec:baryon_tracers}, confirming that FRB sightlines passing through regions of higher projected radio emission and higher radio source density carry systematically larger DMs. The LoTSS radio sky cross-correlation, $\hat{\xi}^{S_\nu \mathcal{D}}(\theta)$ yields per-DM-cut measurements of $1.4\sigma,~2.7\sigma$, and $0.9\sigma$ for the same DM thresholds, combining to a joint SNR of $3.0\sigma$. The LoTSS radio source cross-correlation, $\hat{\xi}^{s_R \mathcal{D}}(\theta)$, similarly yields measurements with per-DM-cut SNRs of $2.0\sigma,~1.6\sigma$, and $1.5\sigma$, combining to a joint SNR of $2.4\sigma$. The broad agreement of the measured signal with theoretical expectations supports the interpretation that these signals arise from the correlated distribution of star-forming galaxies and AGNs with the ionized gas probed by FRB DMs.

\section{Discussion} \label{sec:discussion}

The cross-correlations presented in Section~\ref{sec:measured_correlation_functions} collectively demonstrate that CHIME/FRB DMs carry statistically significant imprints of LSS. In this section, we discuss the broader implications of these measurements. We begin by situating our measurements within the landscape of existing FRB two-point statistic measurements (Section~\ref{subsec:landscape_of_existing_measurements}). We then examine the implications of tSZ and SXRB cross-correlation signal amplitudes for baryonic feedback strength (Section~\ref{subsec:feedback_implications}), and the potential of measuring cluster gas mass fractions (Section~\ref{subsec:baryons_in_clusters_superclusters}), breaking the kSZ effect optical depth degeneracy (Section~\ref{subsec:FRB_kSZ}), and exploring the CGM dust content with CIB correlation (Section~\ref{subsec:cosmic_dust}). We conclude with a discussion of the prospects for this program as FRB samples scale with next-generation facilities (Section~\ref{subsec:future}).

\subsection{Our measurements in the landscape of existing FRB DM two-point statistic measurements} \label{subsec:landscape_of_existing_measurements}

The cross-correlations presented in this work sit within a rapidly maturing observational landscape. Prior to this study, the field had produced a set of pioneering FRB two-point measurements with galaxies and tSZ effect. \citet{2025arXiv250608932W} measured the galaxy$\times$DM angular cross-power spectrum between CHIME/FRB Catalog 2 and DESI BGS across five tomography bins spanning $0.05 < z < 0.5$, achieving a $5.1\sigma$ measurement; the first definitive detection of spatial correlation of FRB DM from plasma surrounding galaxies, implying a characteristic cutoff scale of $\sim 8~h^{-1}$Mpc, a signature of feedback-driven gas suppression. \citet{2025ApJ...993L..27H} stacked the DMs of 61 localized FRBs on photometric galaxy catalogs (WISE-PS1-STRM and DESI), reporting a positive correlation between galaxy number density and DM excess. \citet{2026arXiv260121336S} cross-correlated 2MASS galaxies with 133 localized FRBs, measuring the signal at $\sim 2\sigma$ significance; they interpreted their measurement as evidence for gas mass fractions lower than the global cosmic baryon fraction by at least $\sim 10$\% in group-scale halos. \citet{2025arXiv251102155T} presented the first $4\sigma$ detection of angular cross-correlation between DM excess and Compton-$y$ maps from Planck and ACT, where the measured amplitude implied an average electron temperature of $\sim 2 \times 10^7$~K.

Our work substantially extends this landscape across two dimensions. First, in breadth: we present the broadest suite of FRB DM cross-correlations attempted to date, spanning LSS tracers, thermal and ionized gas probes, demonstrating that even at currently accessible large angular scales, DM correlates robustly with the full range of tracers of cosmic structure. Second, in physical interpretation: as we show in Section~\ref{subsec:feedback_implications}, by leveraging the two correlations most sensitive to feedback -- tSZ$\times$DM and SXRB$\times$DM -- we constrain feedback strength, finding that a weak feedback scenario is disfavored at $\sim 3.5\sigma$ by the SXRB signal, while a fiducial model with moderate feedback is consistent with both probes. Our measurement significance of galaxy$\times$DM is comparable to \citet{2025arXiv250608932W}, a reassuring convergence, given the different galaxy samples, and statistical frameworks employed. The comparable tSZ$\times$DM SNR achieved by \citet{2025arXiv251102155T}, despite using an order-of-magnitude smaller sample of FRBs with arcsecond-scale localizations and host associations, emphasizes the importance of an accurate mean DM subtraction and the ability to probe small scales enabled by precise localizations. Together, our results establish a multi-tracer foundation for future precision studies, demonstrating that the DM field carries detectable astrophysical information across the full baryon landscape.

\subsection{Astrophysical Feedback Strength} \label{subsec:feedback_implications}

Among the suite of cross-correlations presented in this work, the Planck tSZ and RASS SXRB measurements are uniquely sensitive to the strength of baryonic feedback at $\gtrsim 10$~arcmin scales probed here. Since DM traces the line-of-sight integrated electron density ($n_\mathrm{e}$), crossing with X-ray emission ($\propto n_\mathrm{e}^2 f(T_\mathrm{e})$) or the tSZ Compton-$y$ parameter ($\propto n_\mathrm{e}T_\mathrm{e}$) yields stronger weighting towards hot, dense, bound gas, whose distribution is directly sculpted by feedback. These correlations therefore carry sensitivity to feedback, even on the large scales probed here. We measured tSZ$\times$DM and SXRB$\times$DM at SNRs of $3.8\sigma$ and $4.1\sigma$, respectively (Section~\ref{subsec:baryon_tracers}). 

\begin{figure}[ht!]
\centering
\includegraphics[width=\columnwidth]{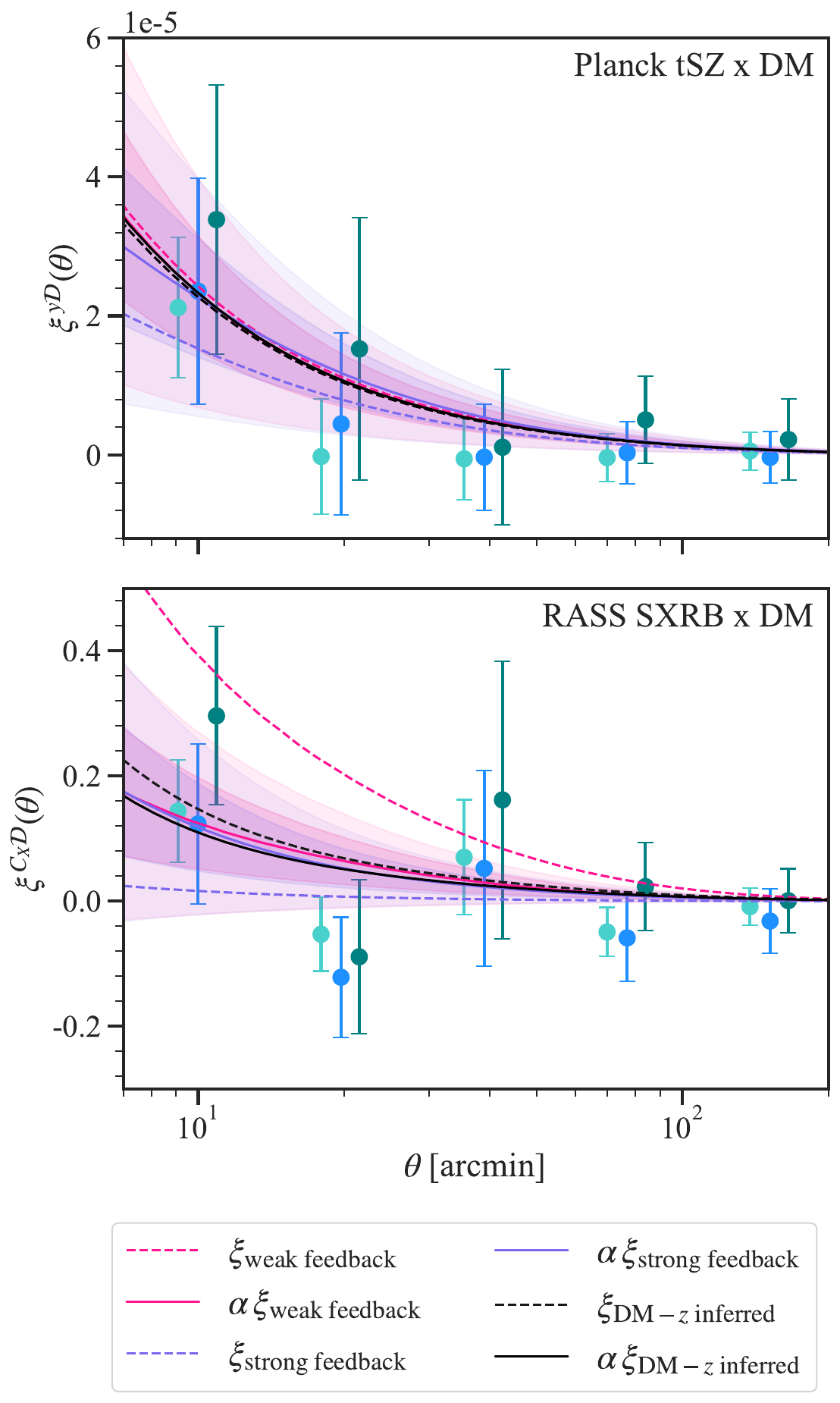}
\caption{Amplitude parameter $\alpha$ fits to the Planck tSZ$\times$DM (top panel) and RASS SXRB$\times$DM (bottom panel), for three feedback scenarios: weak feedback ($\log M_c = 12$, pink), strong feedback ($\log M_c = 14$, blue), and the DM-$z$ inferred feedback model ($\log M_c = 12.84$, black). In each panel, dashed curves show the fiducial theory prediction ($\alpha = 1$) and solid curves show the best-fit amplitude-scaled theory curve ($\alpha\,\xi_\mathrm{th}$), with shaded bands indicating $1/2\sigma$ uncertainty on $\alpha$. For the tSZ measurement, the weak, strong, and DM-$z$ inferred models yield $\alpha = 1.0 \pm 0.3$, $1.5 \pm 0.6$, and $1.0 \pm 0.4$, respectively. For the SXRB measurement, the corresponding values are $\alpha = 0.3 \pm 0.2$, $7.2 \pm 4.2$, and $0.8 \pm 0.4$, respectively. The DM-$z$ inferred model provides a good description of both measurements, while the weak feedback model over-predicts the SXRB signal by $3.5\sigma$.}
\label{fig:feedback_alpha_fits}
\end{figure}

To quantify the consistency of our measurements with different feedback scenarios, we fit an amplitude parameter $\alpha$ to each measurement, defined as the ratio of the observed signal to the theoretical prediction under a given feedback model. Specifically, for a data vector $\hat{\boldsymbol{\xi}}$ and a theory prediction $\boldsymbol{\xi}_\mathrm{th}$, we fit $\alpha$ by minimizing the $\chi^2$,
\begin{equation}
    \chi^2(\alpha) = \left( \hat{\boldsymbol{\xi}} - \alpha \boldsymbol{\xi}_\mathrm{th} \right)^T \hat{\mathbf{C}}^{-1}_\ast \left( \hat{\boldsymbol{\xi}} - \alpha \boldsymbol{\xi}_\mathrm{th} \right),
    \label{eqn:chisq_alpha}
\end{equation}
where $\hat{\mathbf{C}}^{-1}_\ast$ is the Hartlap-corrected inverse covariance matrix (Equation~\ref{eqn:Hartlap_factor}). This yields the analytic maximum-likelihood estimator
\begin{equation}
    \alpha = \dfrac{\boldsymbol{\xi}_\mathrm{th}^T \hat{\mathbf{C}}^{-1}_\ast \hat{\boldsymbol{\xi}}}{\boldsymbol{\xi}_\mathrm{th}^T \hat{\mathbf{C}}^{-1}_\ast \boldsymbol{\xi}_\mathrm{th}}, \, \sigma_\alpha = \dfrac{1}{\sqrt{\boldsymbol{\xi}_\mathrm{th}^T \hat{\mathbf{C}}^{-1}_\ast \boldsymbol{\xi}_\mathrm{th}}}.
    \label{eqn:alpha}
\end{equation}
We evaluate $\alpha$ against two feedback scenarios bracketing the plausible range of baryon physics: a weak feedback model ($\log M_c = 12$), in which gas is retained in halos down to low masses and the signal is maximized, and a strong feedback model ($\log M_c = 14$), in which gas is expelled from all but the most massive halos, substantially suppressing the signal. These bracket the fiducial DM-$z$ inferred model ($\log M_c = 12.84$), which represents our current best estimate of the feedback strength from independent inference~\citep{2026arXiv260417162S}. We illustrate the $\alpha$ measurements in Figure~\ref{fig:feedback_alpha_fits}. 

For the tSZ cross-correlation, we measure $\alpha = 1.0 \pm 0.3$ for the weak feedback model, $\alpha = 1.5 \pm 0.6$ for the strong feedback model, and $\alpha = 1.0 \pm 0.4$ for the fiducial DM-$z$ inferred model. At current noise levels, neither of the two extreme (weak and strong) feedback models is strongly disfavored. For the SXRB cross-correlation, we measure $\alpha = 0.3 \pm 0.2$ for the weak feedback model, $\alpha = 7.2 \pm 4.2$ for the strong feedback model, and $\alpha = 0.8 \pm 0.4$ for the fiducial DM-$z$ inferred model. The weak feedback model substantially over-predicts the observed signal, and the measured amplitude implies a $\sim 3.5\sigma$ tension. The fiducial DM-$z$ inferred model falls between these two extremes and is consistent with both, tSZ and SXRB measurements, providing a reassuring agreement with the feedback constraints derived from a completely independent FRB DM-$z$ relation inference~\citep{2026arXiv260417162S}.

Several sources of uncertainty currently limit the precision of these feedback constraints. Specifically, the imprecision in the DM weighting kernel, arising from the absence of an accurate redshift distribution, compounded by large localization uncertainties, introduces non-trivial systematics that ultimately limit the precision of our conclusions. For the tSZ measurement, the dominant source of uncertainty arises from contamination by the CIB ~\citep{2020ApJ...902...56C, 2025MNRAS.540.1055E, 2025PhRvD.112h3561L}, which is known to correlate with the IGM (Section~\ref{subsec:LSS_tracers}), and may correlate with the host galaxy DM contribution too, introducing an additional layer of uncertainty. 

In the case of the SXRB measurement, several distinct systematics come into play. A well-recognized source of uncertainty is contamination from both resolved and unresolved AGNs, with the precise level depending sensitively on assumptions about the AGN bolometric correction and the intrinsic scatter in the black hole luminosity-halo mass relation~\citep{2024arXiv241212081L, 2026arXiv260202484M}. X-ray binaries constitute another significant contributor to the X-ray sky; given that both X-ray binaries~\citep{2015MNRAS.451.1892A} and FRB progenitors~\citep{2023ApJ...954...80G, 2024Natur.635...61S} are closely linked to the cosmic star-formation history, their correlated contribution to the observed signal is difficult to quantify. A further non-trivial systematic in SXRB analyses arises from the gas metallicity profile assumed to compute the X-ray cooling function (Equation~\ref{eqn:count_rate_profile}). While many studies adopt a uniform metallicity of $0.3~Z_\odot$ for diffuse hot gas~\citep{2024PhRvL.133e1001F, 2023MNRAS.518.1496L}, we instead used empirical profiles from \citet{2018SSRv..214..129M}. Despite this more realistic treatment, metallicity-related uncertainties remain poorly constrained: at group and cluster-scale temperatures ($T \gtrsim 10^6$~K), variations in metallicity from 0 to $Z_\odot$ can induce 20-50\% changes in the cooling function.

These systematic limitations will be substantially reduced in the near future. Samples of arcsecond-scale localized FRBs from next-generation facilities will sharpen the DM weighting kernel. The higher sensitivity, multi-frequency CMB experiments, such as the Simons Observatory~\citep{2024ApJS..274...33G}, will enable improved component separation. The better angular resolution and depth of eROSITA will enable more complete masking of AGNs and X-ray binaries~\citep{2026arXiv260202484M}. Complementary X-ray spectroscopy programs will provide spatially resolved metallicity measurements.

\subsection{Baryons in Galaxy Clusters and Superclusters} \label{subsec:baryons_in_clusters_superclusters}

The DM cross-correlations with X-ray-selected galaxy clusters of median mass $\sim 10^{14.3}~M_\odot$ at $z \sim 0.2$ yield $3.2\sigma$ and $5\sigma$ measurements from eRASS1 and RASS, with mean DM excess of $157 \pm 144$ and $66 \pm 59$~pc\,cm$^{-3}$ at projected separations of $\sim 1-2$~Mpc (see Section~\ref{subsec:baryon_tracers}). These statistical measurements, averaged over hundreds of sightlines, complement the sightline-resolved measurements of DM excess due to ICM. \citet{2023ApJ...949L..26C} reported the discovery of FRB 20220914A and FRB 20220509G, localized by the Deep Synoptic Array to member galaxies of $\sim 10^{14.2}~M_\odot$ Abell clusters, with projected offsets of $\sim$520 and $\sim$870~kpc, respectively, at $z \sim 0.1$. The total observed DMs of $\sim$630 and $\sim$270~pc\,cm$^{-3}$ substantially exceeded the expected mean IGM contributions of $\sim$100 and $\sim$80~pc\,cm$^{-3}$, with the residual excess statistically consistent with IllustrisTNG predictions for the ICM at those impact parameters~\citep{2025arXiv250707090K}. The signal amplitudes measured here at $\sim$1-2 Mpc scales are in broad agreement with their baryonic model.

Extending to the largest gravitationally bound structures, the eRASS1 supercluster cross-correlation yields a $3.3\sigma$ measurement, with mean DM excess of $128 \pm 56$~pc\,cm$^{-3}$ at projected separation of $\sim 2$~Mpc from $\sim 10^{14.7}~M_\odot$ complexes at redshifts $z \sim 0.2$. This measurement is consistent with the direct, map-level evidence for baryon overdensities associated with superclusters recently reported by \citet{2026arXiv260405093R}, who used the same FRB catalog to construct a sky map of extragalactic DM variations across the northern sky. They detected a $\gtrsim 4\sigma$ excess of $\sim 150$~pc\,cm$^{-3}$ extended over $\sim 30^\circ$ scales coincident with the Ursa Major supercluster ($\sim 10^{15.4}~M_\odot$) at $z \sim 0.06$, and a tentative $\sim 2.3\sigma$ excess of comparable amplitude coincident with the Perseus-Pisces supercluster ($\sim 10^{16.1}~M_\odot$) at $z \sim 0.02$. The consistency between these two independent approaches (morphological/map-based vs angular cross-correlation) lends support to the interpretation that the measured DM signals genuinely trace baryons in and around the largest gravitationally bound galaxy cluster complexes in the local universe.

The gas mass fractions implied by the measured DM profiles around galaxy clusters can offer an unbiased calibration of feedback physics in hydrodynamical simulations. Crucially, FRB DMs probe the entire column density of ionized gas along the line of sight, agnostic to its temperature, density phase, or metallicity~\citep{2025ApJ...993..162Z, 2025arXiv250707090K}. This stands in contrast to conventional calibration measurements from X-ray observations, which are strongly weighted towards the densest, hottest, and most metal-enriched gas, and therefore probe only a limited subset of the baryonic content while remaining sensitive to assumptions about metallicity, and thermodynamic state. As large samples of arcsecond-localized FRBs become available, it will be possible to construct statistical DM profiles of galaxy clusters and groups across a wide range of halo masses and redshifts, directly constraining the total gas mass retained within and expelled from halos. Free from systematic uncertainties that affect X-rays, these measurements will provide a clean, model-independent calibration target for hydrodynamical simulations.

\subsection{FRB DM and kSZ Effect Synergy} \label{subsec:FRB_kSZ}

We measured a DM excess of $\sim 10$~pc\,cm $^{-3}$ at $\sim$Mpc separations for the LRG halo population, which is consistent in magnitude with sightline resolved measurements of \citet{2025ApJ...991L..25L} and \citet{2026arXiv260216781M}. A particularly promising avenue opened by the $2.8\sigma$ and $5.1\sigma$ measurements of galaxy$\times$DM correlation with DESI LRGs (this work) and DESI BGS~\citep{2025arXiv250608932W} is the combination of FRB DMs with CMB secondary anisotropies.  

The kSZ measurements have already enabled robust measurements of the projected gas distribution (optical depth) at $\lesssim 2$~Mpc-scales~\citep{2025PhRvD.111b3534H, 2025PhRvD.112h3509H, 2025PhRvD.112j3512R}, and the halo gas mass fractions \citep{2025PhRvD.112l3507H} for these galaxy group samples. These kSZ measurements are highly complementary to FRBs: while the current FRB measurements provide greater sensitivity to large-scale, diffuse gas, kSZ measurements are more powerful on small scales, such that their combination offers a complete picture of the baryon distribution.
Moreover, \citet{2019PhRvD.100j3532M} showed that FRBs can break the optical depth degeneracy that limits the accuracy of large-scale velocity reconstruction from kSZ effect~\citep{2018arXiv181013423S}.  As next-generation CMB experiments, such as the Simons Observatory, deliver kSZ measurements with lower contamination from foregrounds at small scales~\citep{2019PhRvD.100j3532M} and better beam resolution~\citep{2025MNRAS.536..572D}, joint inferences with FRBs to break the optical depth degeneracy will enable percent-level constraints on primordial non-Gaussianity, cosmic growth rate, and amplitude of density fluctuations~\citep{2025arXiv250621657H, 2026arXiv260404867C, 2019PhRvD.100h3508M}.

\subsection{Cosmic Dust in the CGM} \label{subsec:cosmic_dust}

The $4.0\sigma$ CIB$\times$DM signal is particularly compelling from an astrophysical standpoint, as it opens a window onto one of the more contested questions in galaxy evolution: how much dust resides outside galactic disks, in the CGM and beyond~\citep{2010MNRAS.405.1025M, 2024ApJ...974...81R, 2024MNRAS.528.5008O}. The cosmic dust density parameter, $\Omega_\mathrm{dust}$, is a key component of the cosmic baryon inventory~\citep{2004ApJ...616..643F}, and constraining how it is spatially distributed between disks, halos, and the diffuse IGM, directly informs our understanding of stellar feedback, dust grain survival, and metal transport. The CIB captures aggregate far-infrared thermal emission from dust correlated with LSS, including contributions from faint galaxies, diffuse intra-halo emission, and any CGM dust that would otherwise fall below the surface-brightness threshold of galaxy surveys. The observational case for substantial CGM dust was established by \citet{2010MNRAS.405.1025M}, who detected reddening of background quasars extending to several Mpc from foreground SDSS galaxies, inferring a halo dust mass comparable to that in galactic disks. 

Cross-correlating CIB with DM, a diffuse gas tracer, probes the dust-weighted density field on scales bridging halos and the cosmic web. Our theoretical predictions use the $b_\mathrm{CIB}(z) \mathrm{d}I_\nu/\mathrm{d}z$ kernels from \citet{2025ApJ...992...65C}, who performed tomographic cross-correlations of 11 far-infrared maps from Planck, Herschel, and IRAS with SDSS galaxies to reconstruct the evolving CIB spectrum and constrain $\Omega_\mathrm{dust}(z)$. The agreement of our measured CIB$\times$DM amplitude with fiducial theoretical predictions at $0.7\sigma$-level ($\alpha = 0.8 \pm 0.3$) indicates that our measurement is consistent with the cosmic dust evolution peaking at $z \sim 1-1.5$, declining three-folds to the present, and spatial clustering of dusty star-forming galaxies, as characterized by \citet{2025ApJ...992...65C}, 

These measurements will become considerably more informative with a larger, better-localized FRB sample, combined with higher-resolution CIB data. The Planck/IRAS angular resolution ($\sim 5$~arcmin) limits our measurement to scales well above typical virial radii, while the sub-arcminute resolution of, e.g., Herschel~\citep{2010A&A...518L...1P}, the TolTEC camera on the Large Millimeter-wave Telescope~\citep{2018SPIE10708E..0JB}, and the Fred Young Submillimeter Telescope~\citep[formerly CCAT-prime;][]{2023ApJS..264....7C} could reach halo scales. If CGM dust is substantial, the CIB$\times$DM correlation should persist to small separations; a downturn or decorrelation at CGM scales instead would indicate that dust is predominantly confined to star-forming disks, with only a small fraction surviving thermal sputtering in the hot medium, even if galactic winds efficiently launch dust into the CGM. The full-sky spatial-spectral data cube of SPHEREx~\citep{2026ApJ...999..139B} could further isolate the cosmic 3.3~$\mu$m PAH background, tracing very small dust grains and molecules heated stochastically. These small grains may exhibit different survival properties and shielding conditions in the hot halo gas, directly probed by DM, compared to the larger grains traced by far-infrared emission. Together, these complementary probes can build a concordant picture of where cosmic dust resides and how it cycles between interstellar medium (ISM) and CGM.

\subsection{The future of FRB DM $n\times2$-point analyses} \label{subsec:future}

The cross-correlations presented in this work represent an early demonstration of a rapidly maturing observational program enabled by next-generation facilities: the Bustling Universe Radio Survey Telescope in Taiwan~\citep[BURSTT;][]{2022PASP..134i4106L}, the Canadian Hydrogen Observatory and Radio-transient Detector~\citep[CHORD;][]{2019clrp.2020...28V}, the Deep Synoptic Array~\citep[DSA;][]{2019BAAS...51g.255H}, and the Square Kilometer Array~\citep[SKA][]{2004NewAR..48..979C}. As the sample grows from $\sim 3,000$ sightlines with arcminute-scale localizations to $\sim 10^4$ arcsecond-scale localized bursts, the SNR of each individual cross-correlation presented here will improve dramatically (we are currently shot noise dominated).

The statistical forecasts of \citet{2023arXiv230909766R}, \citet{2026ApJ...998..109S}, \citet{2026arXiv260417162S}, and \citet{2026MNRAS.547ag557W} quantify the potential of this program within a joint $3\times2$ and $6\times2$-point framework, combining FRB DM cross-correlations with galaxy clustering, galaxy-galaxy lensing, and cosmic shear from surveys such as the Vera Rubin Observatory~\citep{2019ApJ...873..111I} and Euclid~\citep{2011arXiv1110.3193L}. Incorporating FRB-based 2-point statistics into the standard $3\times2$-point data vector substantially tightens constraints on baryonic feedback: the uncertainty on the characteristic feedback mass scale $\log M_c$ improves by a factor of $\sim 4$ relative to weak lensing alone, propagating into improved constraints on cosmological parameters, including the neutrino mass sum and the dark energy equation-of-state parameters~\citep{2026arXiv260417162S}. The ability to directly measure the feedback-induced suppression of the matter power spectrum will help break the degeneracy between feedback and cosmology that currently limits weak lensing analyses~\citep{2025ApJ...989...81S, 2025arXiv250717742R, 2026arXiv260417162S}.

\section{Conclusion} \label{sec:conclusion}

We have presented the broadest suite of angular cross-correlations of FRB DMs, correlating 3,455 sightlines from CHIME/FRB Catalog 2 with ten tracers spanning the full baryon landscape -- from the dense intracluster medium to diffuse filamentary gas in the cosmic web. Statistically significant measurements at $2.6-5\sigma$ are reported across DESI LRGs, DECADE weak lensing, WISE-reconstructed 100~$\mu$m CIB intensity, Planck CMB lensing and tSZ effect, RASS and eRASS1 X-ray clusters and superclusters, RASS soft X-ray background, and LoTSS radio continuum  sources -- each physically well-motivated and broadly consistent with theoretical predictions. Together, these constitute the most comprehensive multi-tracer characterization of the FRB DM perturbation field to date.

Beyond establishing statistical measurements, these cross-correlations carry quantitative astrophysical content. The tSZ and SXRB cross-correlations constrain baryonic feedback strength, with the DM-$z$ inferred model providing a consistent description of both probes while a weak feedback scenario is disfavored at $\sim 3.5\sigma$ by the SXRB signal. The statistically significant measurement of large-scale DM excess around galaxies, groups, and clusters opens the door to joint analyses with kSZ effect measurements at small scales to address the long-standing optical depth degeneracy in kSZ velocity reconstruction, and establishes that gas fraction measurements from projected DM profiles may serve as an unbiased calibrator for simulations in future. The correlation with CIB reveals the promise for dissecting the dust content between galactic disk and CGM at halo-scales. The correlations with radio sources may pave one of the only ways to probe baryonic physics at high redshifts.

The systematic limitations of this work -- large localization uncertainties and imprecise DM weighting kernel arising from unknown FRB redshifts -- are not fundamental barriers. In the near-term, we can target FRB tomography using cross-correlations with galaxies in several redshift bins~\citep{2020PhRvD.102b3528R} and CIB observed at various frequencies~\citep{2025ApJ...992...65C}. As next generation facilities (BURSTT, CHORD, DSA, and SKA) deliver samples of $10^4$ arcsecond-scale localized FRBs, each of these systematics will be addressed, and the signal-to-noise of every cross-correlation presented here will improve dramatically. In this regime, FRB two-point statistics will transition from measurement to inference, providing direct constraints on feedback physics and cosmological parameters alike. The present work is both, a proof of concept and a quantitative foundation for a new era in which FRBs serve as precision cosmological \textit{backlights}, illuminating the full baryon content of the cosmic web.

\begin{figure*}[ht!]
\centering
\includegraphics[width=\textwidth]{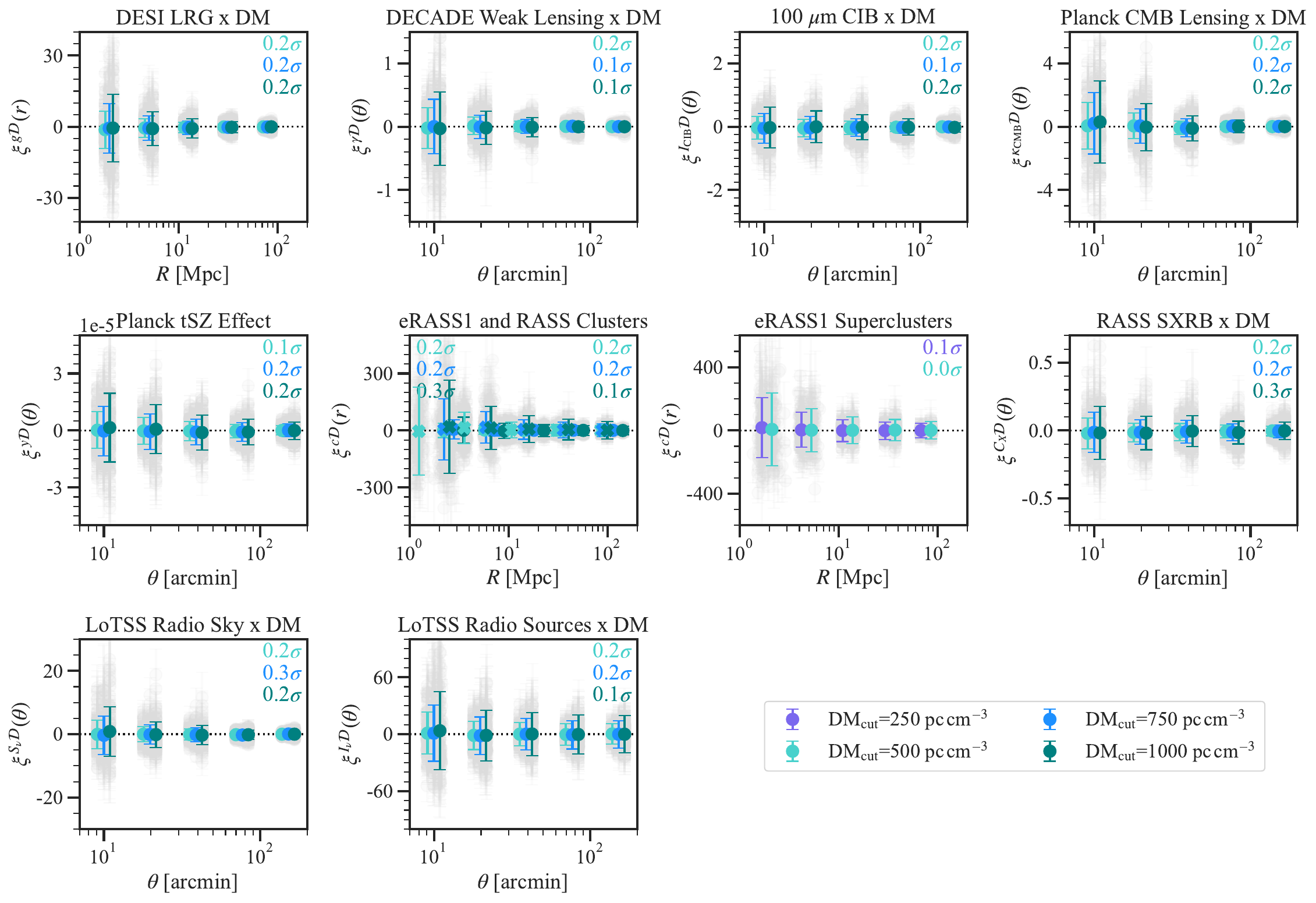}
\caption{Null test validation of correlation functions between CHIME/FRB DMs and each of the ten tracers, measured with jackknife covariance estimator over 50 spatial patches. The DMs of FRBs were randomly shuffled to construct the null expectation. Gray points in the background represents measurement from each of the 100 shuffles. The colored points denote their mean, and the uncertainties include the shot noise and the uncertainty on mean estimate.}
\label{fig:null_tests_summary_onlyDMshuffled}
\end{figure*}

\section*{Acknowledgments}
During the preparation of this work, KS, EK, WW, SF, and YC were supported in part by grant NSF PHY-2309135 to the Kavli Institute for Theoretical Physics (KITP). This material is based upon work supported in part by the National Science Foundation under CAREER Grant Number 2240032, and the David \& Lucile Packard Foundation grant 2020-71384. SG acknowledges support provided by the Austrian Research Promotion Agency (FFG) and the Federal Ministry of the Republic of Austria for Climate Action, Environment, Energy, Mobility, Innovation and Technology (BMK) via the Austrian Space Applications Programme with grant numbers 899537, 900565, and 91197. We thank Cameron Hummels, Carlos Garcia-Garcia, Daisuke Nagai, Elena Pierpaoli, Gilbert Holder, Jiachuan Xu, Joop Schaye, Kaitlyn Shin, Kendrick Smith, Martijn Oei, Matthew Johnson, Matthew Madhavacheril, Neal Dalal, Nicholas Battaglia, Selim Hotinli, Stefano Borgani, and Tassia Ferreira for insightful discussions. We gratefully acknowledge the CHIME/FRB Collaboration for producing the largest FRB catalog published to date that forms the foundation of this work.

\facilities{
CHIME, 
Dark Energy Spectroscopic Instrument, 
Dark Energy Camera, 
Wide-field Infrared Survey Explorer, Planck, 
ROSAT, 
eROSITA, 
LOFAR, 
Infrared Astronomical Satellite, 
COBE.
}

\software{
\textsc{Astropy},
\textsc{APEC},
\textsc{BaryonForge}, 
\textsc{Cartopy},
\textsc{Healpy},
\textsc{Pyccl},
\textsc{SciPy},
\textsc{TreeCorr}.
}

\appendix

\section{Null Expectation Validation}\label{sec:null_tests}

We validate our measurements by conducting null tests to quantify systematic effects arising from the small sample size, which could otherwise produce fluctuations that mimic a spurious signal. We randomly shuffle the FRB DMs while keeping their on-sky positions fixed to de-correlate the baryon distribution with the LSS or baryon tracers to construct the null expectation. The results from 100 shuffles are presented in Figure~\ref{fig:null_tests_summary_onlyDMshuffled} and the corresponding SNR$_\mathrm{null}$ of the signal averaged over these 100 shuffles are listed in Table~\ref{table:detection_statistics}. These measurements are consistent with the de-correlation/null expectation.

\bibliography{manuscript}{}
\bibliographystyle{aasjournal}

\end{document}